\documentclass[preprint]{aastex}
\usepackage{natbib}

\slugcomment{}

\shorttitle{Exoplanets Detection Around Multiple Starsi}
\shortauthors{Thomas et al.}

\begin{document}


\title{Techniques for High Contrast Imaging in Multi-Star Systems I: Super-Nyquist Wavefront Control}


\author{S. Thomas\altaffilmark{1} and R. Belikov\altaffilmark{1}  and E. Bendek\altaffilmark{1} }
\affil{NASA AMES Research Center, Moffett Field, CA 94035, USA}
\begin{abstract}
Extra-solar planets direct imaging is now a reality with the deployment and commissioning of the first generation of specialized ground-based instruments (GPI, SPHERE, P1640 and SCExAO). These systems allow of planets $10{^7}$ times fainter than their host star.  
For space-based missions (EXCEDE, EXO-C, EXO-S, WFIRST),  various teams have demonstrated laboratory contrasts reaching $10^{-10}$ within a few diffraction limits from the star. However, all of these current and future systems are designed to detect faint planets around a single host star or unresolved multiples, while most non M-dwarf stars such as Alpha Centauri  belong to multi-star systems.  Direct imaging around binaries/multiple systems at a level of contrast allowing Earth-like planet detection is challenging because the region of interest is contaminated by the hostÕs star companion as well as the host Generally, the light leakage is caused by both diffraction and aberrations in the system. Moreover, the region of interest usually falls outside the correcting zone of the deformable mirror (DM) for the companion.
Until now, it has been thought that removing the light of a companion star is too challenging, leading to the exclusion of binary systems from target lists of direct imaging coronographic missions. 

In this paper, we will show different new techniques for high-contrast imaging of planets around multi-star systems and detail the Super-Nyquist Wavefront Control (SNWC) method, which allows to control wavefront errors beyond nominal control region of the DM.  Using the SNWC we reached contrasts around $5\times10^{-9}$ in a 10\% bandwidth. 
\end{abstract}



\section{Introduction}

\begin{figure}[h]
\begin{center}
\includegraphics[scale=0.3]{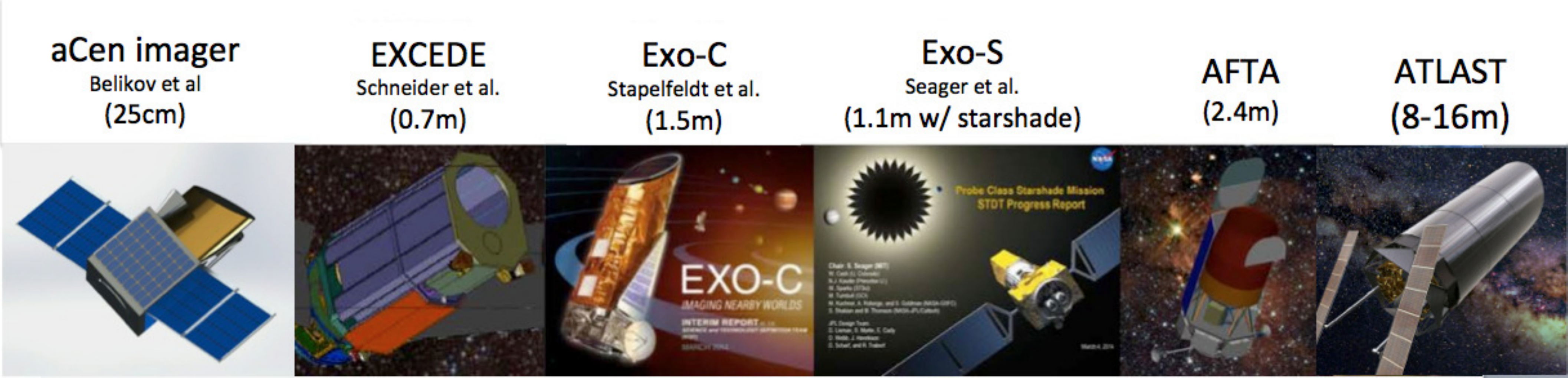}\\
\caption{\label{fig:missions} Mission concepts compatible with the methods proposed here. For these missions, no hardware changes are required as long as they have a deformable mirror. }
\end{center}
\end{figure}

The exoplanets field is rapidly expanding with the success of the Kepler mission (\citep{Burke14} and references therein) and the emergence of direct imaging ground based instruments (GPI \citep{Macintosh14}, SPHERE \citep{Beuzit08}, SCExAO \citep{Guyon10}, P1640 \citep{Hinkley08}. One of the most exciting prospects of future telescopes is finding other Earths analogues in our galaxy or solar neighborhood and more ambitiously detect life on them. The Kepler space telescope has already revealed that roughly 22\% of stars have planets between 1 and 2 Earth radii in their habitable zone \citep{Batalha14}.
However, this mission does not perform spectral characterization of these targets. Direct imaging combined with spectroscopic characterization of exo-Earths would allow us to determine the chemical composition of the planet's atmosphere, and constrain the presence of oxygen, water and other elements necessary for life. Over the past decade, there have been more than a dozen direct imaging mission studies for space-bases telescopes. Figure \ref{fig:missions} shows a few representative missions into which a lot of these have evolved. 

An Earth-like planet orbiting the habitable zone of a sun-like star would be 10 billion times dimmer than the star.  Diffraction created by the telescope aperture as well as aberrations is several orders of magnitude brighter than the planet, making its detection very challenging. Several starlight suppression systems \citep{Guyon10b, Kern13} have been demonstrated $10^{-9}$ raw contrast or better in the laboratory.

These systems employ high performance coronagraphs to suppress the star diffraction created by the telescope aperture and efficient wavefront control systems based on a deformable mirror (DM) to remove residual starlight leak (speckles) created by the imperfections of telescope optics. However, their designs have thus far been mostly limited  to single-star systems and to planets or disks found within the DM control zone, where speckles can be corrected by conventional means. Binary star systems are good candidates to search for planets since they are more common than single stars. Good examples of such binary systems are Alpha Centauri ($\alpha$ Cen) and Sirius. Currently, these multi-star systems are typically excluded from mission target lists because there is no technical approach that can manage 
 the technical challenges associated with double-star (or multi-star) high-contrast imaging. The three main challenges are:\\
- The first challenge is that often the multi-star separation is typically beyond the half-Nyquist frequency of the deformable mirror. In this paper, we propose a  new method called ÒSuper-Nyquist Wavefront ControlÓ (SNWC) that uses a mild grating (or an existing pattern, print-through, commonly found on many Deformable Mirrors left over from their manufacturing process) to effectively alias low-spatial frequency modes of the DM into higher frequencies, enabling the DM to remove speckles well beyond the DM's Nyquist frequency. In effect, aliasing is used as a feature rather than a bug.\\
- The second challenge is to separate and independently remove overlapping speckles from multiple stars. The solution is called Multi-Star Wavefront Control (MSWC). We solve this challenge by selecting a region of interest such that different DM modes are used for the two different stars (at the expense of reducing the discovery region). 
 \\
- The third challenge  is to create a dark zone for a multi-star system where the planet angular separation star requires SNWC to control diffraction in the discovery region, and the companion star creates diffraction and aberration light in the same discovery zone. In this case we combined the SNWC and MSWC together, and call this a technique Multi-Star Super Nyquist Wavefront Control (MSSNWC). 

Each of these techniques serve  specific science case that we will present in section \ref{sec:sciencecase}. In this paper, we focus on the Super Nyquist algorithm. After discussing the challenges of observing multi-star systems in section \ref{sec:challenges}, we present the theoretical background of the method in section \ref{sec:SNMSWC_all}. We also briefly explain the multi-star wavefront correction. However the details of the MSWC is beyond the scope of the paper. Finally, we show simulation results in section \ref{sec:simul} both in monochromatic and polychromatic light. 

\section{Science motivation}
\label{sec:sciencecase}
\subsection{Searching for planets around multi-star systems}
The science cases addressed by several exoplanet detection missions such as the ones shown in Figure  \ref{fig:missions} may result in great leaps in our understanding of warm disks, exoplanet diversity, dynamics, and atmospheres. They also will deliver a census of exoplanets around nearby stars, and (for some missions) the detection and spectral characterization of Earth- like planets in the habitable zone. However, none of these missions are planning to image multi-star systems with current technology, except systems for which the leak and glare of the companion star is negligible. 

\begin{figure}[h]
\begin{center}
\includegraphics[scale=1.5]{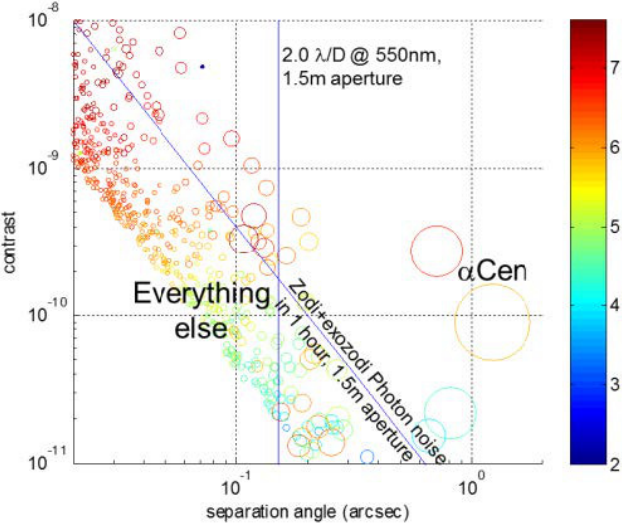}
 \caption{Simulation showing an Earth twin at maximum separation around every real nearby star. On this contrast vs separation angle graph, the circle size and
color represent star size and type. $\alpha$Cen is ~3 times easier than any other star by almost any metric, except for the fact that it is a binary. The vertical line also shows the particular case of where 2 $\lambda/d$ would be placed for a 1.5m aperture telescope at 550nm.}
\label{fig:AlphaCen} 
\end{center}
\end{figure}

The method presented in this paper will greatly multiply the science yield of these missions since it enables direct imaging of planetary systems around multi-star systems, without additional hardware changes or costs. Enabling the study of multi-star systems is very important because the majority of K-type and earlier stars are in multi-star systems. In particular, 5 out of the 7 star systems within 4 parsecs containing K- or earlier type stars are multiples ($\alpha$ Cen, Sirius, Procyon, 61 Cyg, $\epsilon$ Ind), and only two are singles ($\epsilon$ Eri, $\tau$ Cet). While it is true that there are many more nearby M-dwarfs and most of them are isolated, the direct imaging of M dwarfs is arguably better done from the ground with Extremely Large Telescopes (ELTs) because: (a) they are dimmer and require larger apertures; (b) their planetary systems are closer to the star and require the angular resolution of larger apertures; 
Conversely, the study of K- and earlier type stars arguably favors space-based missions, for which the smaller apertures are less concerning, but require much deeper contrasts only possible from space. Therefore, most of the stars best suited for space missions are in fact in multi-star systems and SNWC combined or not with MSWC promise to greatly increase the science yield of almost any direct imaging mission, as well as enable the study of a whole new class of star systems, namely multiple stars. 
One case deserves special attention. SNWC and MSWC enable high contrast imaging around our nearest-neighbor star, $\alpha$Cen. 
High-contrast imaging with an inner working angles of 2 $\lambda/d$ are now realistic, a 30cm telescope should be sufficient to image the habitable zone at high-contrast. The $\alpha$Cen system represents a particularly favorable target for planet detection missions, but it has been excluded from current missions' target lists because of its multiplicity.
Stars of comparable proximity to $\alpha$ Cen are all very dim, and stars of comparable brightness are about three times farther away. In particular, the next closest star earlier than M-type ($\epsilon$ Eri) is 2.4 times as far, and is known to have a thick disk that may interfere with detection of small planets. The next star of comparable proximity to  $\alpha$ Cen is BarnardÕs star, which is 1.4 times farther, has a much dimmer magnitude (M=10), and has a habitable zone only 30 mas wide, requiring at least a 4m aperture to even resolve it (Figure \ref{fig:AlphaCen}).

Another reason why $\alpha$ Cen is an attractive target is because recent estimates of $\eta_{\Earth}$ from Kepler have been on the order of 10\% per octave of semi-major axis and per octave of planet size, leading to integrated numbers of ~10- 22\% under typical assumptions of habitable zone size and habitable planet size range \citep{Batalha14}. Therefore $\alpha$ Cen has about a 20-40\% chance of harboring an exo-Earth around either the A or B star.
A mission using the techniques proposed here, may be the first to detect and spectrally characterize an Earth twin, if one exists around $\alpha$ Cen  \citep{Dumusque12}.

\subsection{Exoplanetary debris systems}

\begin{figure}[h]
\begin{center}
\includegraphics[scale=1]{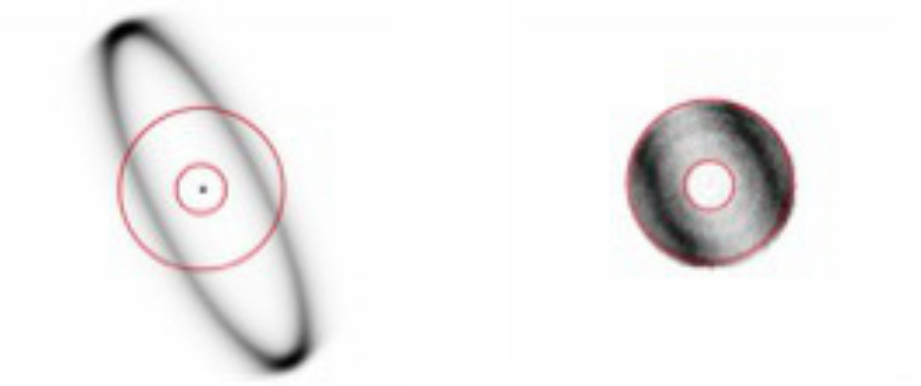}\\
\caption{\label{fig:disks} Simulations (courtesy of Thomas P. Greene, private communication) showing that for the example of HR4796A, the disk is truncated with the current capabilities of the AFTA instrument.The simulations were done using the Zodipic package, a general-purpose modeling tool for optically-thin disks \citep{Kuchner12}. 
}
\end{center}
\end{figure}

SNWC allows another specific science case. When using SNWC, the outer working angle (OWA) and thus the field of view (FoV) of most missions in Figure \ref{fig:missions} is limited by the DM half Nyquist frequency. As an example, for 2.4m telescope with a 48x48 DM such as AFTA configuration, this implies an outer working angle of about 1" (depending on wavelength). Thus, the disk around $\epsilon$ Eri can only be seen in high contrast out to $\sim$4AU. For the case of $\alpha$Cen, AFTA would not be able to observe beyond 1AU even if the second star was not there. The SNWC method resolves this outer working angle limitation since it enables the extension of the dark zone for single-star systems past the DM Nyquist limit. SNWC does not increase the size of the dark zone, but it does allow a user to shift it to arbitrary locations. By stitching separately acquired sub- and super-Nyquist dark regions, imaging of arbitrarily large disks is enabled.
For instance, simulations \footnote{courtesy of Thomas P. Greene, private communication}, show that HR4796A, the disk is truncated with the current capabilities of the AFTA coronagraph instrument (Figure \ref{fig:disks}). 

\section{Challenges of multi-star and extended disks observing}
\label{sec:challenges}
 As mentioned in the introduction, coronagraphy is challenging in the context of multiple-star systems because even in the case of external coronagraphy, the companion star leaks light into the region of interest around the target. Moreover, the separations between the two stars is such that it is outside the correctable field of view (FoV) set by the number of degrees of freedom of the deformable mirror. In the context of extended disks, the limitation is the limited FoV.
 \begin{figure}[h]
\begin{center}
\includegraphics[scale=0.3]{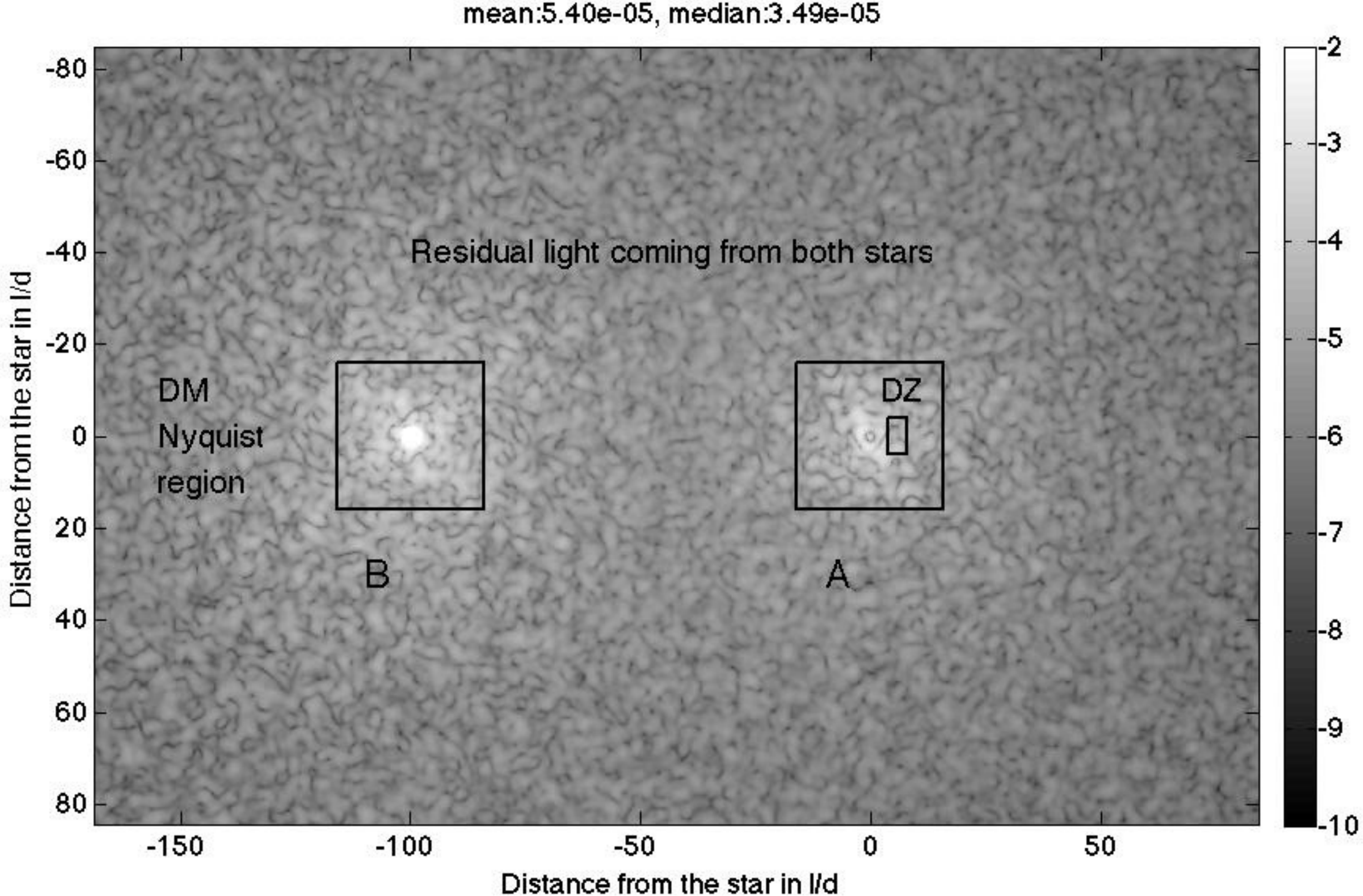}
 \caption{Schematic showing the limitations of the multi-star coronagraphy. The image shows overlapping speckle fields from the two companions (A and B). The black squares show the standard control zone of the deformable mirror with respect to each star, which we refer to as the "sub-Nyquist region" (of that star) and the rectangle called DZ is the dark zone region, where we want to detect a planet. The mean and median contrast of this dark zone are respectively 5.4 $\times 10^{-5}$ and 3.49 $\times 10^{-5}$} 
\label{fig:SNMSWC_chalenges} 
\end{center}
\end{figure}
\pagebreak
\paragraph{Light leakage from the companion star(s)}
When observing a target belonging to a binary system, the amount of light leaking from the off-axis companion (hereafter referred to as the companion) reduces the Signal to Noise Ratio (SNR) of the candidate planet around the on-axis target (hereafter referred to as the target) to a level that the planet candidate might become undetectable. Two fundamentally different effects contribute to creating this light leak: diffraction and optical aberrations. These effects are complicated by the fact that the beams coming from the two stars are incoherent from each other.
 \begin{figure}[h]
\begin{center}
\begin{tabular}{ccc}
\includegraphics[scale=0.18]{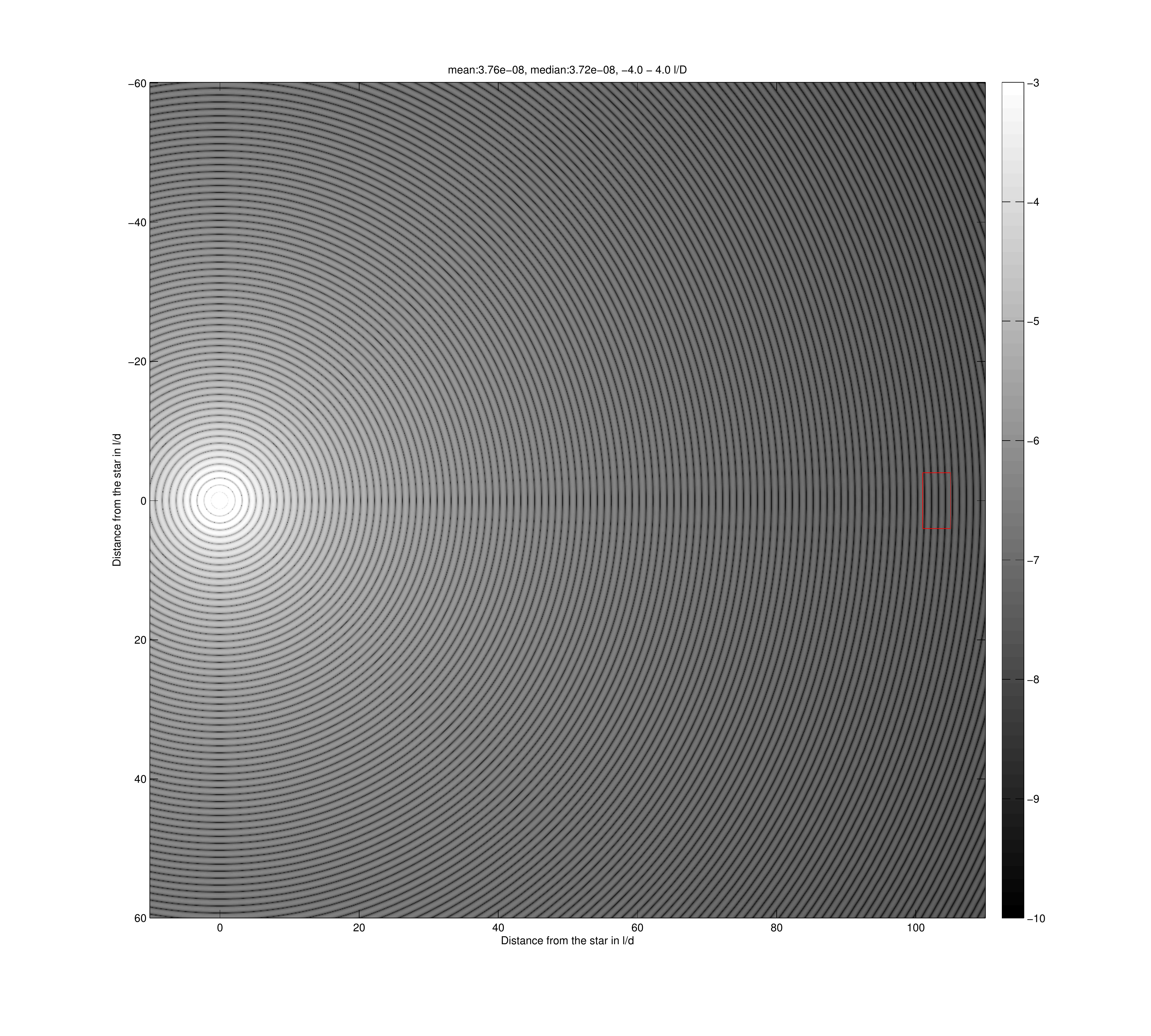}&\includegraphics[scale=0.18]{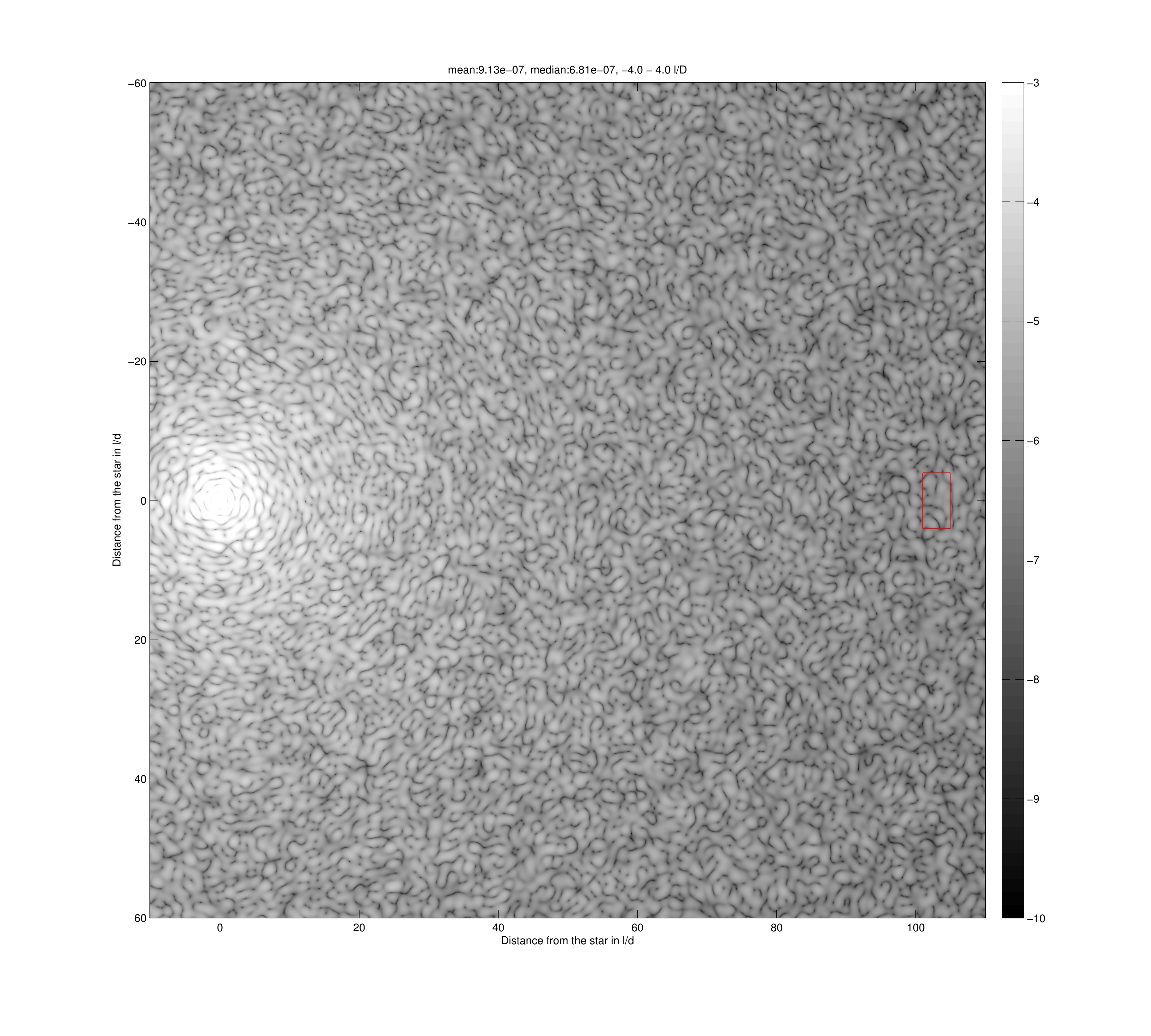}\\
\includegraphics[scale=0.4]{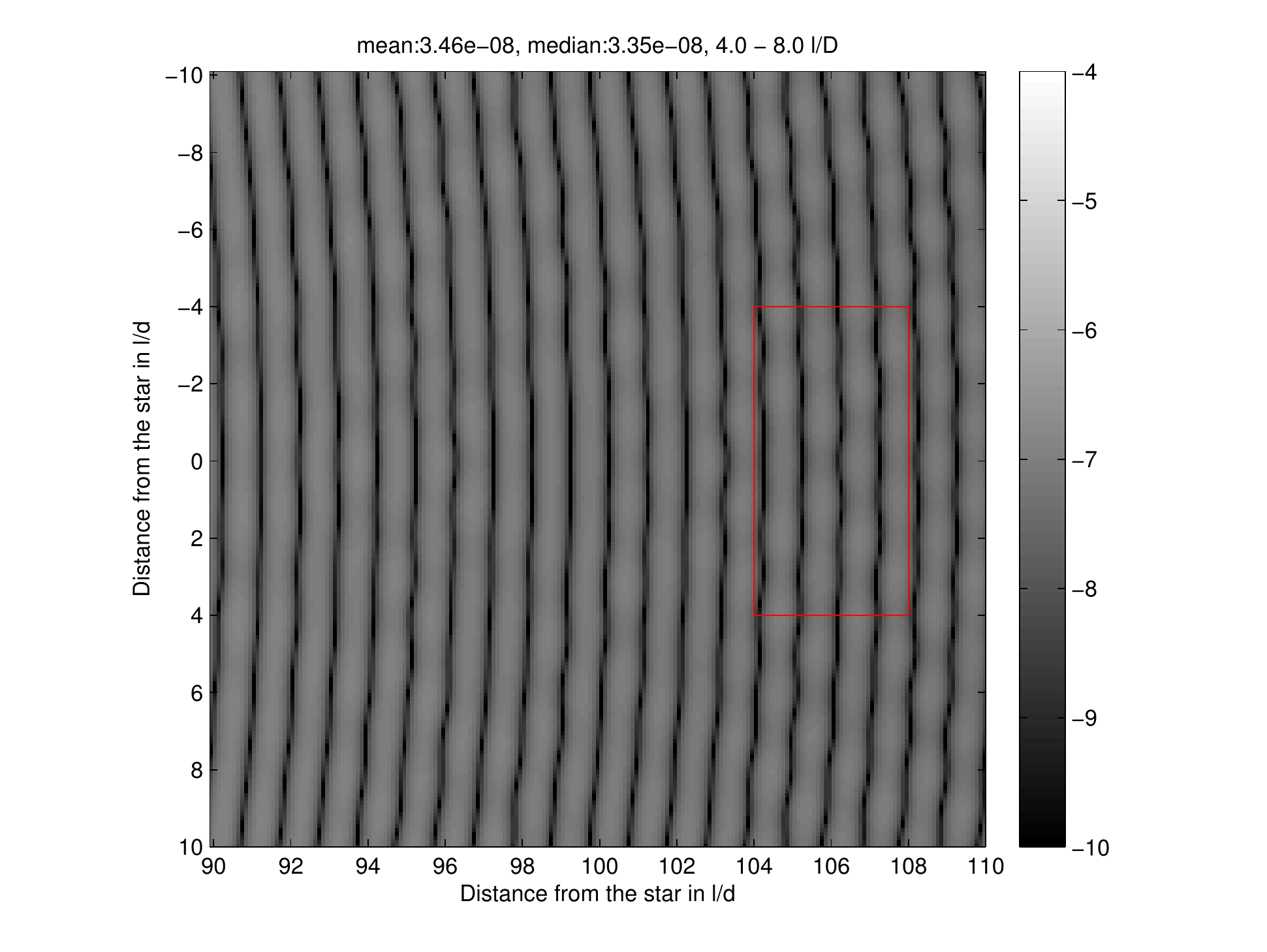}& \includegraphics[scale=0.4]{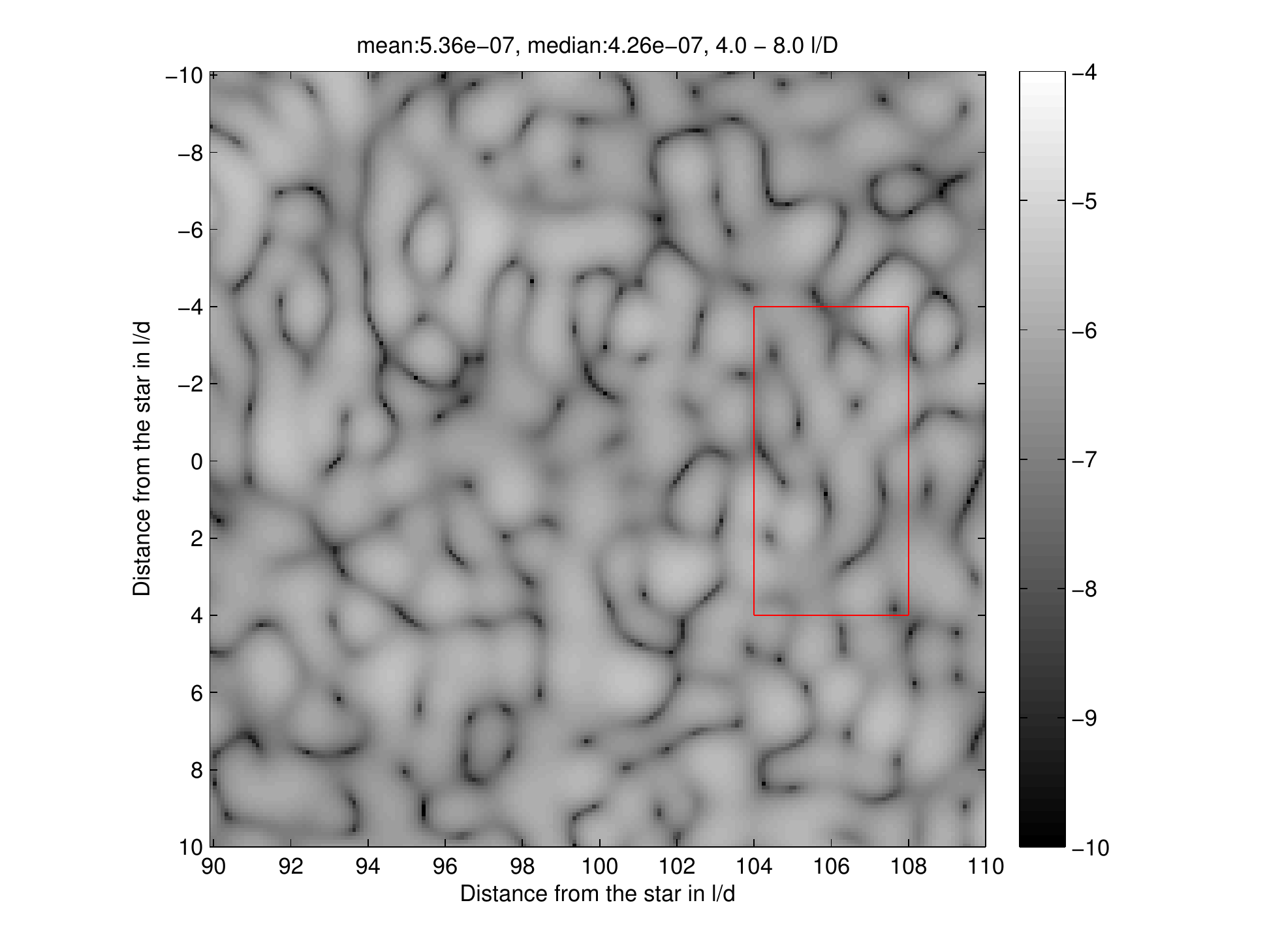}\\
\end{tabular}
\caption{Light diffracted at 100 $\lambda/D$ in monochromatic light, without aberrations (left) and with 25nm rms aberrations (right). The size of the region of interest is 4$\times$8 $\lambda/D$. The top views are larger field of views showing the companion star. The bottom is a zoom on the region of interest (close to the suppressed on axis star). The median contrast intensity of the order of 4$\times$ 10$^{-8}$ without aberrations and 4$\times$ 10$^{-7}$ with 25nm rms of aberrations. The contrast is shown in a perfect scenario (meaning totally removing the light from the on-axis target).}
\label{fig:DiffLight}
\end{center}
\end{figure}

To quantify the light leakage, we consider the case of searching for a planet or a disk located at 100 $\lambda$/d away from a target star. This could be a binary separated by 10" observed through a 1.5m telescope at a wavelength of 770 nm or a single star observed with a large telescope at shorter wavelength.Figure \ref{fig:DiffLight} shows the light diffracted by the companion in the particular scenario where the on-axis star is totally suppressed (valid in the context of EXO-S for instance, for which the star shade practically removes all the light from the target star). The simulated contrast is  the intensity at a given position in the image normalized by the peak intensity of the target. The simulation was done in the context of no aberrations (left image) and with 25nm rms of aberrations with a power law of $f^{-2}$ (right image). The figures also show the dark zone region (DZ) as a box. We use these regions as a reference for the initial median contrast before corrections.

\begin{figure}[h]
\begin{center}
 \includegraphics[scale=0.6]{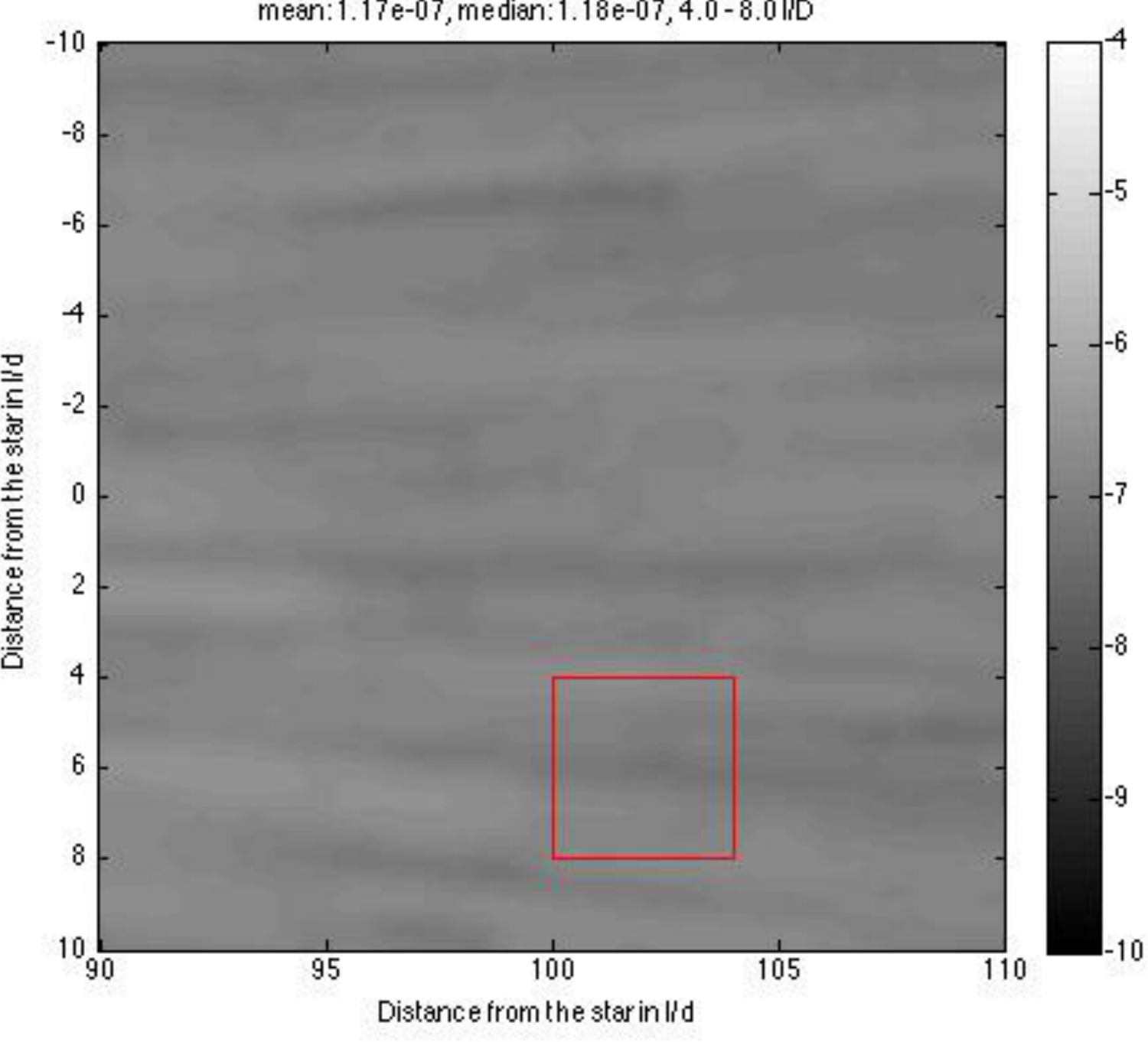}
\caption{ Light diffracted past 100 $\lambda/D$ away from a star in polychromatic light (10\% bandwidth), with 10nm rms aberrations. The size of the region of interest is 4$\times$4 $\lambda/D$. The median contrast intensity is of the order of 1e$^{-7}$ with 10nm rms of aberrations, masking the light coming from a potential planet. }
\label{fig:DiffLightPoly}
\end{center}
\end{figure}

 The amount of light originating from the companion depends on the aberrations amplitude and their spatial frequency distribution in the system. In monochromatic light and in presence of no aberrations, the amount of light diffracted in the DZ is of the order of 4e$^{-8}$.  With 25nm rms of aberration, the contrast worsens to 4e$^{-7}$.
 Figure \ref{fig:DiffLightPoly} shows the polychromatic case, for a 10\% bandwidth around the central wavelength of 550nm. The median contrast intensity is of the order of 2e$^{-8}$ without aberrations and 2e$^{-7}$ with 10nm rms of aberrations.   In monochromatic light the diffraction rings are blurred together and therefore not shown on Figure \ref{fig:DiffLightPoly}.

Until now the standard approach to control diffracted light from the second star in a binary system has been to design a coronagraph to block both stars. A coronagraph is only useful if the diffraction dominates over aberrations or otherwise cannot be removed by the wavefront control system. As a general rule, for typical mirror aberrations, diffraction dominates close to the star but aberrations dominate far from the star. 
All coronagraphs suppress diffraction, but they do not help with aberrations. A wavefront control (WC) system (or equivalent) is required to suppress optical aberrations, and in addition to suppressing aberrations, is often capable of suppressing diffraction by at least an order of magnitude. Therefore, a coronagraph is neither sufficient nor necessary to suppress the leak of the off-axis star, while a wavefront control system is both necessary and often sufficient to suppress both the aberrations and diffraction.

\begin{figure}[h]
\begin{center}
\includegraphics[scale=0.6]{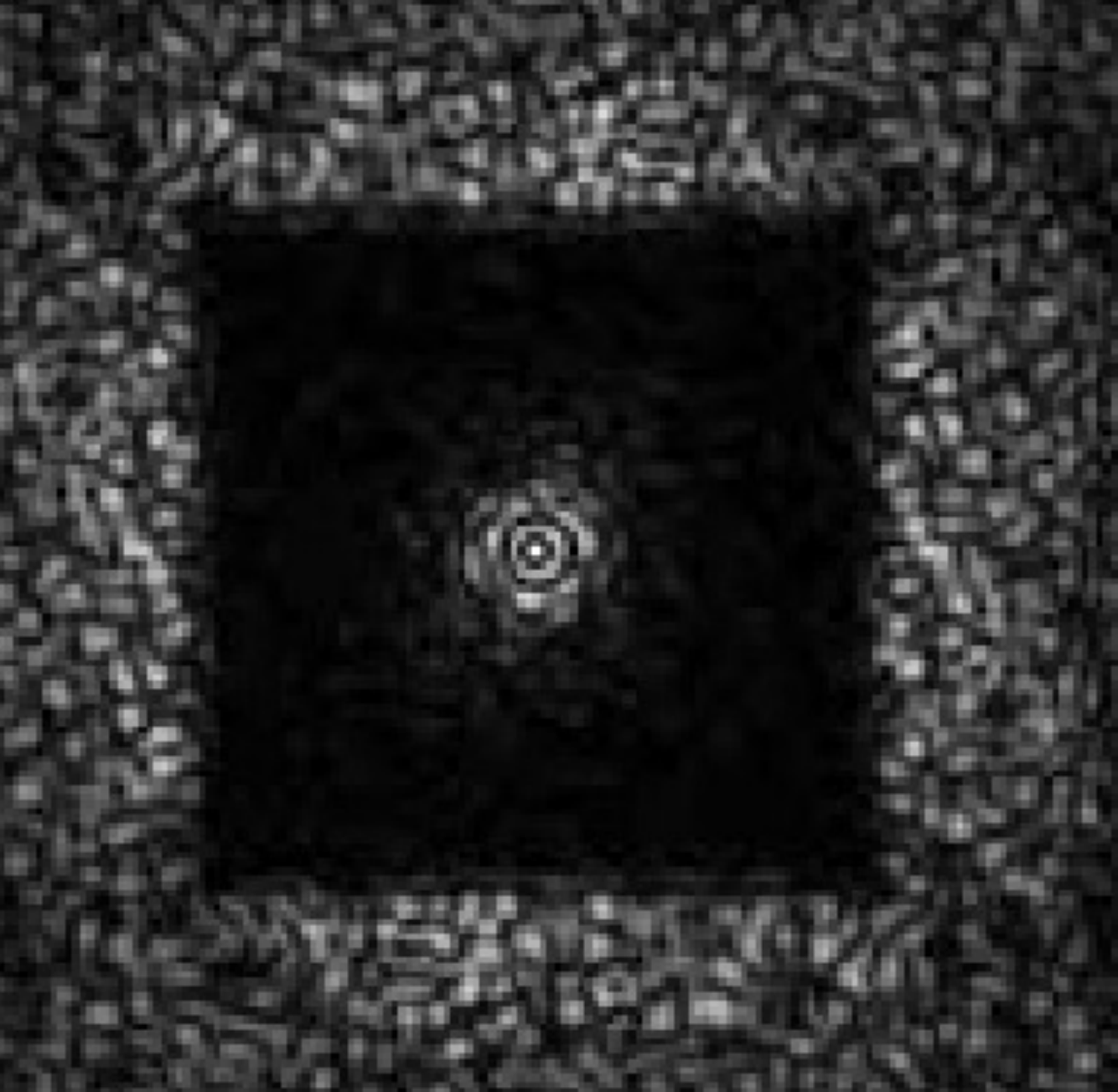}
 \caption{Simulation of a dark zone with a discovery region limited by the number of actuators on the deformable mirror. This is a case where only phase is corrected leading to a symmetrical dark zone. The coronagraph blocks most of the light before an image plan wavefront control algorithm further removes speckles. }
\label{fig:DMlimitations} 
\end{center}
\end{figure}

\paragraph{Sub-Nyquist region of the DM}
The nominal region over which we can create a dark zone is defined by the number of actuators on the deformable mirror. Figure \ref{fig:DMlimitations} illustrates this issue. The square region around the central star is the controllable region and is limited by the Nyquist frequency $f_N$, which corresponds to an outer working sky angle of to $N_{act} \times \lambda/2D$. It has been previously believed that the deformable mirror can not control any speckles past this Nyquist-limited outer working angle.

\section{Theory of Super-Nyquist Multi-Star Wavefront Control}
\label{sec:SNMSWC_all}

\subsection{Super-Nyquist Wavefront Control}\label{sec:SNWC}

We propose here to develop a solution based on wavefront control rather than using a new coronagraph design. Our solution to the super-Nyquiest limitation uses a mild grating (or an existing pattern commonly found on many DMs left over from their manufacturing process) to remove light coming from the off-axis star.  The grating creates a controlled diffracted image of the off-axis star close to the on-axis star, placing the diffracted off-axis image within the DM control region, enabling it to remove the speckles from both stars (which can be done independently at the cost of halving the size of each control region). This solution uses the DM in a regime outside its nominal control range, where the DM experiences what is typically a harmful side effect: spatial aliasing. We use aliasing as a feature. 

 In this section we present the theoretical foundation for Super-Nyquist Wavefront Control and explore its theoretical capabilities and limitations. We also discuss how to design DM parameters for a particular application. Before we begin, we introduce a simplified formalism for general coronagraphic wavefront control.

Consider some arbitrary coronagraph with a deformable mirror and a science focal plane. Let $E_{DM}(x)$ and $E_f(\xi)$ be the electric fields in those two planes, where $x$ and $\xi$ are 2-vectors representing the normalized 2D coordinates in units of pupil size $D$ and sky angle $\lambda/D$, respectively. Because the coronagraph is a passive linear system, the relationship between these two fields is given by some linear operator $\mathcal{C}$:
\begin{eqnarray}
\label{eq:Ef}
E_f=\mathcal{C}\{E_{DM}\}
\end{eqnarray}

A change in the deformable mirror setting creates a change $\Delta E_{DM}$ in the DM field and a corresponding change $\Delta E_f=\mathcal{C}\{\Delta E_{DM}\}$ in the focal plane.

For purposes of our simplified treatment, we will assume that most coronagraphs can be approximated as follows:
\begin{eqnarray}
\label{eq:C}
\mathcal{C}\{E_{DM}(x)\}(\xi)=\mathcal{F}\{A(x) E_{DM}(x)\}T(\xi)
\end{eqnarray}
where $\mathcal{F}$ is the Fourier Transform (with our normalization of $x$ and $\xi$, the Fraunhofer integral reduces to the Fourier Transform to within a constant phase factor), $A(x)$ represents the coronagraph aperture (or more generally, the cumulative effect of all apertures projected or propagated onto the DM), and $T(\xi)$ represents the coronagraphic throughput as a function of sky angle, normalized to a maximum value of 1. Typically $A(x)=1$ across a large portion of the DM, and $T(\xi)=1$ almost everywhere in the focal plane except the small blind spot with radius of a few $\lambda/D$ corresponding to where the star is suppressed by the coronagraph. In what follows, we make the following simplifying assumptions: (a) we only consider the regions outside the coronagraphic blind spot, which allows us to set $T(\xi)=1$, and (b) the aperture $A(x)$ is binary valued (0 or 1), and is 1 across most of the aperture. Note that with these assumptions and our normalizations, Eq.(\ref{eq:C}) will be in units of contrast. With these assumptions and simplifications, we get:
\begin{eqnarray}
\Delta E_f = \mathcal{F}\{\Delta E_{DM}\}
\end{eqnarray}

The key idea behind SNWC is that a DM can diffract some light in $\Delta E_f$ beyond the sub-Nyquist region into super-Nyquist regions, thus enabling control of those regions. Here, by "sub-Nyquist" we mean the region within the spatial half-Nyquist frequency of the DM (corresponding to 0th diffraction order), and by "super-Nyquist" we mean regions beyond the spatial half-Nyquist frequency of the DM (corresponding to 1st and higher diffraction orders). For example, for a 32x32 actuator DM, the sub-Nyquist region would be within $16 \lambda/D$ outer working angle. Conventional high contrast wavefront control is limited to this sub-Nyquist region only. However, because DMs diffract and modulate light in super-Nyquist regions, stellar leak in those regions can in principle be controlled in essentially the same way as in the sub-Nyquist region. The main difference is that the contrast level of DM modulation will generally be weaker in super-Nyquist regions than in sub-Nyquist because the amount of light diffracted into those regions by the DM is generally small. We will quantify and show how to mitigate this effect as part of the analysis in this section.

Light can be diffracted either due to, (a) the periodic actuator nature of the DM, including any print-through pattern, (b) external gratings (which may not share the actuator periodicity). We treat those separately in the two subsections below.

\subsection{SNWC with diffraction caused by the DM influence function or print-through pattern}
\label{sec:inf}
\begin{figure}[h]
\begin{center}
\includegraphics[scale=0.6, angle=90]{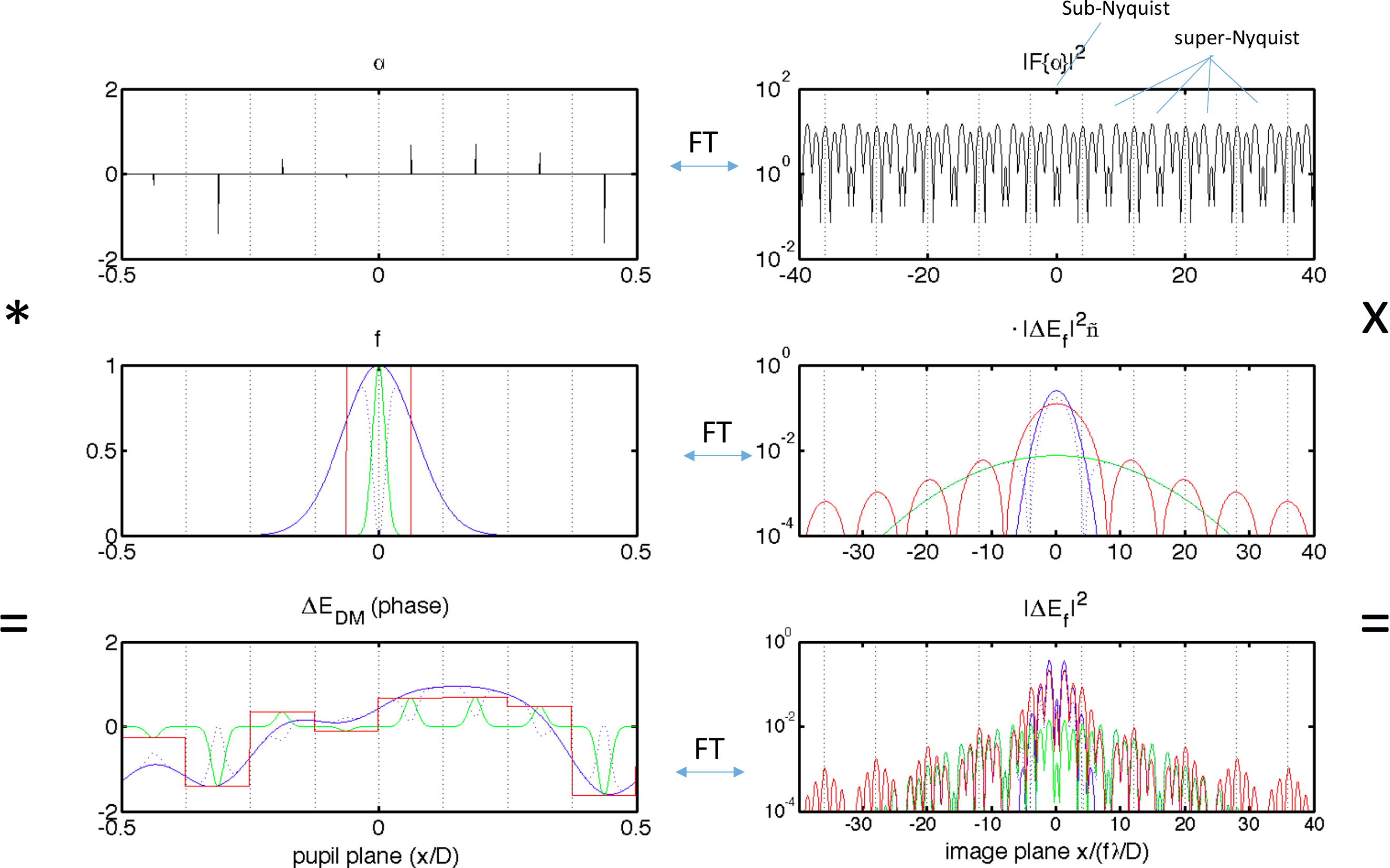}\\
\caption{\label{fig:theoSNWC1} DM print-through case. The left column shows the pupil plane and the right column the image plane. The top row  shows the actuator coefficients $\alpha$ and its Fourier Transform (Fourier series of $a_n$). The second row shows realistic influence functions and the bottom row shows the convolution or multiplication of the top two rows.}
\end{center}
\end{figure}

The DM field is given by:
\begin{eqnarray}
\label{eq:Edm}
E_{DM}(x)
&=& A(x)e^{i\phi_{DM}(x)}\\
\nonumber
&=& A(x) [ 1+i\phi_{DM}(x)+o(\phi_{DM}(x)^2)]
\end{eqnarray}
where $\phi_{DM}(x)$ is the phase imparted to the electric field by the DM in radians. The first term of the above equation corresponds to the on-axis PSF, while the rest is the contribution is due to the DM which we will call $\Delta E_{DM}$. For small DM modulations, we can assume that only the leading term is significant ($i\phi$) and use the influence function model:
\begin{eqnarray}
\nonumber
\Delta E_{DM}(x) &=& i\phi_{DM}(x)\\
&=&i\sum\limits_{n=1}^Na_nf(x-nd)
\end{eqnarray}
where $f$ is the DM influence function, $d$ is the spacing of actuators on the DM, and $a_n$ (for $n=1...N$) are the DM actuator coefficients. We also dropped $A(x)$ for simplicity and will simply assume the equivalent condition that $\phi_{DM}(x) = 0$ outside the DM aperture. We will adopt the convention of $f$ being normalized to unity maximum value, which implies that $a_n$ are in units of radians. We can re-express this equation as a convolution:
\begin{eqnarray}
\nonumber
\Delta E_{DM}(x)=i\alpha \ast f
\label{eq:tconv}
\end{eqnarray}
where $\alpha=\sum\limits_{n=1}^Na_n\delta(x-nd)$. This convolution can be seen on the left column of Figure \ref{fig:theoSNWC1} . We now proceed to analyze the behavior in the focal plane by computing the electric field change created by the DM:
\begin{eqnarray}
\nonumber
\Delta E_f &=& \mathcal{F}\{\Delta E_{DM}\}\\
&=&i\mathcal{F}\{\alpha\}\mathcal{F}\{f\}
\end{eqnarray}
Computing the intensities (which with our normalized units will be in units of contrast) gives:
\begin{eqnarray}
\label{eq:periodic}
|\Delta E_f|^2 = |\mathcal{F}\{\alpha\}|^2|\mathcal{F}\{f\}|^2
\end{eqnarray}
$\mathcal{F}\{\alpha\}$ is a periodic function (see Figure \ref{fig:theoSNWC1} , top row), because the periodic sampling of $\alpha$ by delta functions makes $a_n$ the Fourier Series of $\mathcal{F}{\alpha}$. Its different periods correspond to the sub-Nyquist region periodically copied, or aliased, into super-Nyquist regions. In other words, if influence functions were delta functions (a non-physical hypothetical scenario), any speckle modulations by the DM in the sub-Nyquist region would be perfectly periodically repeated (aliased) into all super-Nyquist regions. In the case of realistic influence functions, the super-Nyquist regions will get attenuated (as we show below), imposing a coupling between the width of the influence function, DM stroke and maximum correctable contrast of errors in super-Nyquist regions. This coupling is not simple to characterize in the general case, but becomes very simple if we characterize it in a statistical sense. We can treat the DM actuator coefficients $a_n$ as independent random variables with standard deviation $\sqrt{\left<a_n^2\right>}$ radians (where without loss of generality we assume a 0 mean). This standard deviation is essentially a measure of DM stroke. For example, in a 32x32 actuator DM, the peak stroke will typically be ~3x this amount. 

Random variations of the DM lead to random variations of the focal plane field $E_f$ ("speckles"). The power spectral density of these variations is:

\begin{eqnarray}
\label{eq:periodic2}
\left<|\Delta E_f|^2\right> &=& \left<|\mathcal{F}\{\alpha\}|^2\right>|\mathcal{F}\{f\}|^2\\
\nonumber
&=&\sum\limits_{n=1}^N\left<a_n^2\right>|\mathcal{F}\{\delta(x-nd)\}|^2|\mathcal{F}\{f\}|^2\\
\nonumber
&=&N\left<a_n^2\right>|\mathcal{F}\{f\}|^2\\
\nonumber
\end{eqnarray}
This power spectral density is essentially a measure of speckle contrast (as a function of position in the image plane) correctable by a DM with stroke $\sqrt{\left<a_n^2\right>}$. As expected, higher strokes lead to higher energy or contrast in the DM correction field and therefore the ability to correct for higher levels of error. Because the vast majority of the DMs are capable of strokes of several radians ($\sqrt{\left<a_n^2\right>}>1$), and because our linearity assumption breaks down for ($\sqrt{\left<a_n^2\right>}>1$), we can assume that for the vast majority of cases,  $\sqrt{\left<a_n^2\right>}=1$, simplifying the above to:
\begin{eqnarray}
\label{eq:final}
\left<|\Delta E_f|^2\right> &=& N|\mathcal{F}\{f\}|^2\\
\nonumber
\end{eqnarray}
which states that the Fourier Transform of the influence function times the number of DM actuators $N$ directly gives a measure of the contrast correctable by the DM, in sub- as well as super-Nyquist regions.

Figure \ref{fig:theoSNWC1}  shows practical applications and implications of this equation. The left column shows the pupil plane and the right column the image plane. The top row  shows the actuator coefficients $\alpha$ and its Fourier Transform (Fourier series of $a_n$). This would be the situation if the influence function was a delta function (unphysical). In this case, the correction from a DM in the focal plane is a periodic function repeating in each Nyquist-sized region. Controlling or modulating speckles in the central sub-Nyquist region would be perfectly repeated in all super-Nyquist regions, enabling independent suppression of starlight in any super-Nyquist region in exactly the same fashion as in the sub-Nyquist region. The second row shows realistic influence functions and the bottom row shows the convolution or multiplication of the top two rows. Note that the middle right plot also represents the contrast of error correctable in super-Nyquist regions per (\ref{eq:final}). We now consider four representative cases of influence functions.

Typical DM influence functions in continuous-sheet DMs (Figure  \ref{fig:theoSNWC1}, middle left, blue) are roughly the size of actuator spacing. The corresponding $\left<|\Delta E_f|^2\right>$ (Figure  \ref{fig:theoSNWC1}, middle right, blue) tells us that there is only significant correction in the sub-Nyquist region while super-Nyquist regions are attenuated too much to be controllable at any reasonable contrast levels. Roughly speaking, because such an influence function is a factor of $N$ smaller than the DM itself (in area for 2D), (\ref{eq:final}) tells us that the width (area) of $\left<|\Delta E_f|^2\right>$ is $N \lambda/D$ ($N (\lambda/D)^2$ for 2D) and the mean value is $1/N$. This simply states the well-known fact that conventional DMs are capable of suppressing starlight leak in the sub-Nyquist region, and that the mean contrast of that leak cannot be higher than ~$1/N$, averaged across the sub-Nyquist region, by energy conservation. For example, in a 32x32 DM ($N = 1024$), the sub-Nyquist region is $1024 (\lambda/D)^2$ in area (32 x 32 $\lambda/D$) and the controllable level of starlight leak is $10^{-3}$ contrast averaged across the sub-Nyquist region. For such a DM, super-Nyquist control requires a grating or print-through pattern on the DM (covered in the next subsection).

Consider now a continuous sheet DM with an unconventional influence function that is much narrower than the actuator spacings (green). Specifically, suppose it is a factor of $M$ smaller than the blue influence function (in area for 2D). Its Fourier Transform intensity $|\mathcal{F}\{f\}|^2$ (Figure \ref{fig:theoSNWC1}  middle right, green) shows its ability to correct in super-Nyquist. It will be a factor of $M$ wider (in area for 2D) and a factor of $M$ dimmer than the case of a conventional DM (blue), so as compared to a conventional DM, it can control super-Nyquist regions, at the expense of lowering the contrast of errors it can control in the sub-Nyquist region. Roughly speaking, such a DM is capable of correcting any one of M super-Nyquist regions, as well as the sub-Nyquist region, as long as the contrast of the errors is not higher than ~$1/MN$, averaged across any one region. Note that we can create such an "unconventional" DM using existing conventional DMs. For example a conventional 64x64 DM with only every 8th row and column connected is effectively an 8x8 DM with undersized influence functions. In this case, $N$=8x8 and $M$=8x8, the DM can still control speckles out to 32 $\lambda/D$, but only one 8x8 $\lambda/D$-sized region at a time. If the number of electrical lines is a cost or risk driver, and a mission emphasis is on planet characterization rather than search (i.e. planet location is known), such a DM may be preferred to the fully connected 64x64 DM.

Now consider the case of a conventional influence function (blue), but with a dip in the middle from print-through manufacturing (blue dotted curve in Figure \ref{fig:theoSNWC1}). This influence function was created simply by taking the difference between the blue and green influence functions. As a result, the corresponding $\left<|\Delta E_f|^2\right>$ has essentially the same controllability of the sub-Nyquist region as the solid blue curve, but also can control super-Nyquist regions as well as the green case. The relationship between the width of the narrow dip in the influence function, the contrast and the number of super-Nyquist regions that can be controlled is the same as for the green case.

Finally, a segmented DM (red color) has both a full control of the sub-Nyquist region (to full contrast levels similar to the conventional continuous-sheet DM case in blue) and can correct super-Nyquist regions, as long as the error does not exceed the contrast shown by the red curve when spatially averaged across any super-Nyquist region. This ability comes from the side lobes of the red curve. Correction will be best at odd multiples of half-Nyquist frequency (peaks of sidelobes) and there will be a blind spot at multiples of Nyquist frequency (zero-crossings of the red curve). In 2 dimensions, these sidelobes will be strong only along directions normal to the atuator edges (along "diffraction spikes" of the PSF).

\subsection{SNWC with a grating or beamsplitter}
\label{sec:grat}
DM diffraction into super-Nyquist regions can be caused not only by a particular shape of the influence function, but also directly by a mild external grating. The case of a grating with periodicity matching the DM actuator periodicity has already been treated in the previous subsection (dotted blue case in Figure \ref{fig:theoSNWC1}), and here we consider the case where the grating has a finer periodicity, which can be designed for a particular desired super-Nyquist distance and contrast level.

In the previous subsection, the action of the diffraction-causing agent (influence function) was a convolution in the pupil plane and a multiplication in the image plane. In this subsection, the action of the diffraction-causing agent (grating) is the opposite: a multiplication in the pupil plane and a convolution in the image plane.

Assume the same scenario as in the previous subsection, except with a mild grating represented by a periodic function $g(x)$ multiplying the DM field (\ref{eq:Edm2}). Define the new DM field as:
\begin{eqnarray}
\label{eq:Edm2}
E_{DM,g}(x)
&=& E_{DM}(x)g(x)\\
\nonumber
&=& A(x)(1 + \Delta E_{DM}(x))g(x)\\
\nonumber
&=& A(x)g(x) + \Delta E_{DM}(x)g(x)
\end{eqnarray}
where $E_{DM}$ is the field from the previous subsection (i.e. without the presence of the grating). In the image plane, the first term will lead to the on-axis star PSF, together with fixed (DM-independent) attenuated copies of the star PSF diffracted by the grating. Each of these PSF copies can be thought of as a new star around which conventional sub-Nyquist wavefront control can be applied, thus enabling super-Nyquist wavefront control with respect to the original star. However, each PSF copy will create a small "blind spot" that the DM will not be able to remove, similar to the blind spot created by the original star in the center of the image. (If the periodicity of the grating is the same as the DM pitch, as would be the case if the grating is the DM print-through pattern, then there will be a small blind spot in the exact middle of every super-Nyquist region.) For purposes of this subsection, we will assume that any light these PSF copies create outside the blind spots are part of the star leak error to be suppressed. Because they are independent of the DM, we do not book-keep them as part of the DM correction field and thus ignore the first term, focusing only on terms that are DM-dependent. The perturbation of the DM electric field by the DM is then:
\begin{eqnarray}
\Delta E_{DM,g}(x)=\Delta E_{DM}(x)g(x)
\end{eqnarray}
Fourier transforming to the image plane leads to the perturbation by the DM of the image plane electric field:
\begin{eqnarray}
\Delta E_{f,g}(\xi)=\Delta E_f(\xi) G(\xi)
\end{eqnarray}
where $\xi$ is the image plane position, $\Delta E_f$ is the image field perturbation from the previous subsection (i.e. without the grating), and $G = \mathcal{F}\{g\}$. Because $g$ is a periodic function, its Fourier Transform has the form: $G=\sum{g_n}\delta(\xi-n\Delta\xi)$ where $g_n$ are the Fourier Series coefficients of $g(x)$ and $\Delta\xi$ is the spacing between the diffraction orders in the focal plane (equal to the spacing between super-Nyquist regions if the grating periodicity is the same as the DM actuator periodicity).
 
\begin{figure}[h]
\begin{center}
\includegraphics[scale=0.7, angle=90]{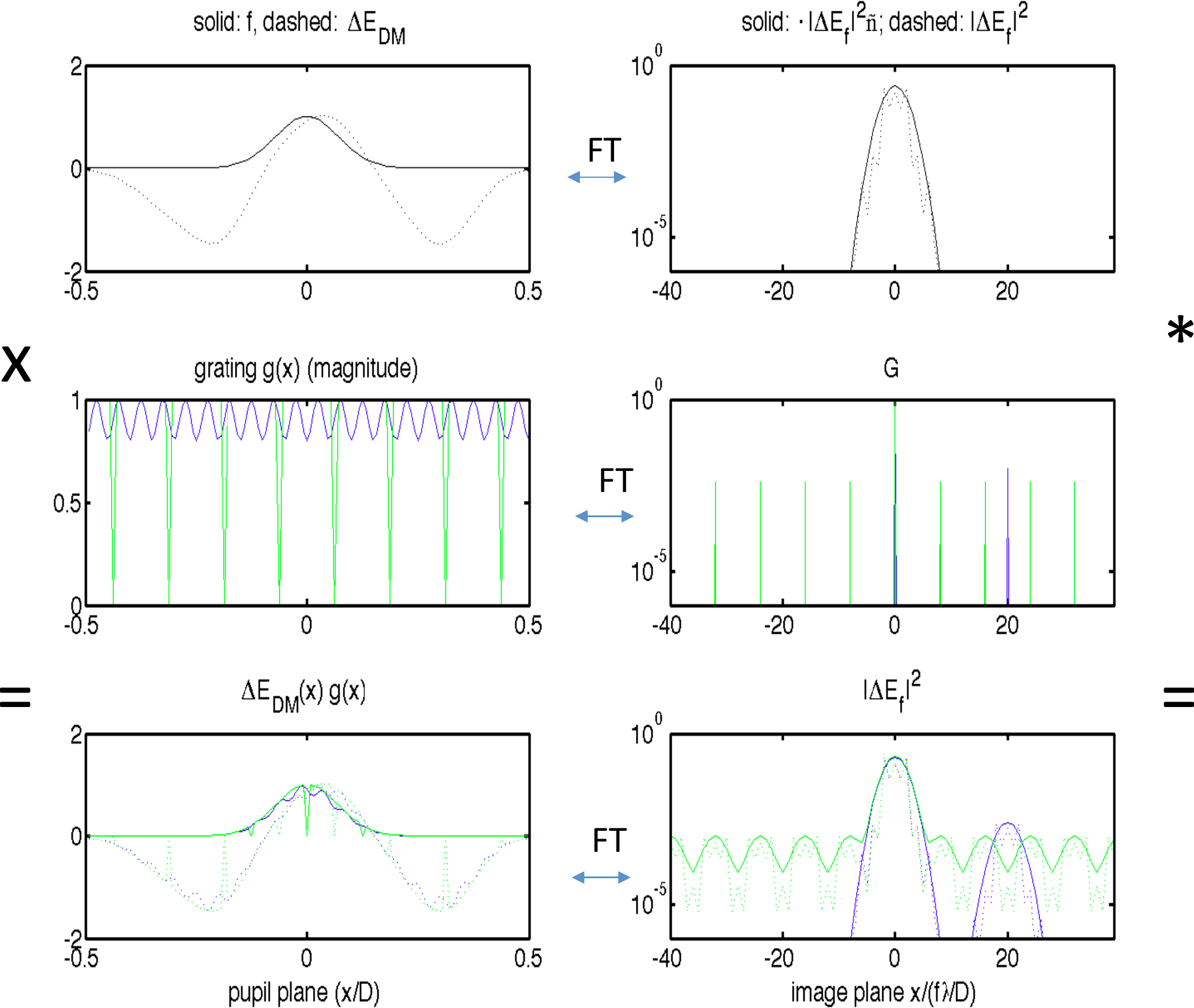}\\
\caption{\label{fig:theoSNWC2} Grating case. The left column shows the pupil plane and the right column the image plane. The top row  shows the DM shape and its Fourier Transform (Fourier series of $a_n$). The second row shows the grid pattern and the bottom row shows the multiplication or convolution of the top two rows. }
\end{center}
\end{figure}

Following the same statistical characterization method as in the previous subsection, we treat the perturbations of the image plane electric field by the DM as a random variable and compute its power spectral density, which after some algebra becomes:
\begin{eqnarray}
\left<|\Delta E_{f,g}|^2\right>=N|\mathcal{F}\{f\}|^2\ast|G|^2
\end{eqnarray}
In other words, the sub-Nyquist control region is copied to the diffraction orders of the grating $g$ and attenuated by a factor equal to the contrast of the diffraction order. Two examples are shown in Figure \ref{fig:theoSNWC2} . The blue case corresponds to the case where $G$ only has one off-axis term, which is simply a mild beamsplitter bleeding off 1\% of the total beam into 20 $\lambda/D$. This is the most efficient way to control a specific fixed super-Nyquist region. The green case in Figure \ref{fig:theoSNWC2}  corresponds to a periodic amplitude grating such as a print-through on the DM. It acts in much the same way as a beamsplitter, except it creates many super-Nyquist control regions instead of just one.

There are a few key principles relating the grating design characteristics and the super-Nyquist control contrast and region location: (a) when a grating creates a diffraction order (PSF copy) of contrast $C$, super-Nyquist control is enabled around that diffraction order and up to $N/2 \lambda/D$ away, in exactly the same fashion as sub-Nyquist control around the original on-axis star; (b) the total energy of the error to be corrected cannot exceed the energy in that diffraction order. For example, if the diffraction order is $10^{-3}$ contrast and we have a 32x32 DM correcting in a 32x16 $\lambda/D$ half-region around that diffraction order, then the average contrast of the error corrected cannot exceed $10^{-3}$/(32x16) = $2\times10^{-6}$. (On the other hand, if the region of interest is only 3x3 $\lambda/D$, then speckles of up to $10^{-4}$ contrast can be corrected.

\subsubsection{SNWC implementation with a grating}

Using dots on the pupil has been proposed previously to calibrate dynamic distortions on wide-field optical systems enabling high-precision astrometric measurements \citep{Guyon12}. For this technique the dots can be arranged in the pupil using a hexagonal geometry allowing higher azimuthal sampling. The diffractive pupil spacing can be adjusted to create PSF replicas to run the SNWC, and also to obtain high-precision astrometry on wide-field images. A description of the optimal hexagonal geometry for combining these two techniques has been published by \citep{Bendek13}.

\subsection{Multi-Star Wavefront Control}
\label{sec:MSWC}
\begin{figure}[h]
\begin{center}
\includegraphics[scale=0.5]{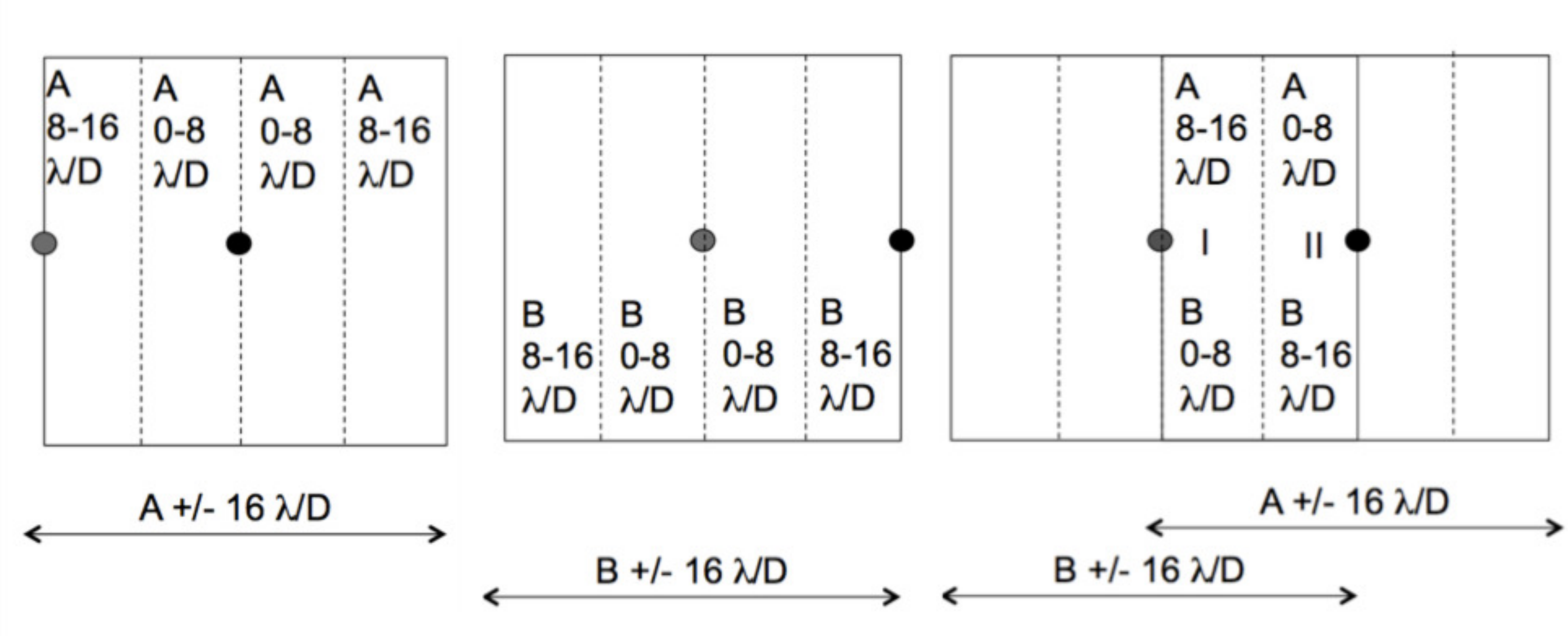}
 \caption{Use of different DM modes to independently modulate light for two different stars. The first scheme on the left shows the different controllable regions for the star A (black dot) and the middle the controllable regions for the star B (gray dot). The figure on the right represents the superposition of the two and shows that there are two regions names I and II between star A and star B for which we can control A and B at the same time with the same DM. For region I we would use the modes from 0 to 8 $\lambda/D$ for B and 8 to 16 $\lambda/D$ for A and vice et versa.}
\label{fig:MSWC} 
\end{center}
\end{figure}

For sake of completeness, we will describe here how to simultaneously suppress the speckle fields of both stars (MSWC). Consider the case of two stars, A and B, with (for now) a sub-Nyquist separation.  The main challenge is that light from the two stars is mutually incoherent, and therefore light from each star can only be used to destructively interfere its own speckle field but not the field from the other star. In order to suppress both stars, it is necessary to be able to independently modulate the speckle field of each star without affecting the speckle field of the other (in some region of interest). Figure \ref{fig:MSWC} shows the special case of two stars separated by 16 $\lambda$/D. The left panel shows the (sub-Nyquist) control region of a 32x32 DM with respect to star A, and separates the region into 4 vertical sections, each controlled by a different and independent set of modes on the DM (the outer sections are controlled by modes on the DM corresponding to spatial frequencies of 8-16 cycles per aperture (cpa), and the inner regions are controlled by 0-8 cpa). The middle panel of Figure \ref{fig:MSWC} shows the same thing, but with respect to star B. Finally, the right panel superimposes these control regions of the two stars. The two regions between the two stars are labeled as ÒIÓ and ÒIIÓ . In region I, the 0-8 cpa modes modulate the speckle field of the B star but not the A star, and the 8-16 cpa modes modulate the speckle field of the B star but not the A star. In other words, in region I, the speckle field of the A star can be suppressed by using the 8-16 cpa modes without affecting the B star and the speckle field of the B star can be independently suppressed by using the 0-8 cpa modes without affecting the A star. In effect, we have reduced the MSWC problem to two separate conventional WC problems (each using different DM modes), which we know how to solve. Solving these two conventional WC problems simultaneously results in simultaneous suppression of the speckle fields of both stars in region I. The same can be done in region II (but not simultaneously with region I). The final result is that a double star dark zone can be created at the expense of reducing the size of the control region by a factor of two (once again, one cannot cheat the number of degrees of freedom available on the DM), and there are two such regions. These two smaller regions can be suppressed separately and then stitched together to create the full field of view between the two stars.

This idea can be generalized to the case of arbitrary (sub-Nyquist) star separation. The result is that one can always partition the intersection of the sub-Nyquist regions of the two stars into two sections, each with half the area of the original control region, where speckle fields of both stars can be simultaneously suppressed. (These regions will have shapes different from Figure \ref{fig:MSWC} and may consist of disconnected parts.) A generalization to N stars, implies N independent correction regions, each with 1/N of the original control region area. These regions are found to be so-called Voronoi partitions of a periodically extended star field.
It should be noted that in practice it is impossible to completely decouple DM modes. Any DM mode will always affect all stars everywhere to some level. However, on the regions we constructed, one of the starÕs speckle fields is affected much more than the other, so the above algorithm works in closed loop.

\subsection{Super-Nyquist Multi-Star Wavefront Control}
\label{sec:SNMSWC}
\begin{figure}[h]
\begin{center}
\includegraphics[scale=0.45]{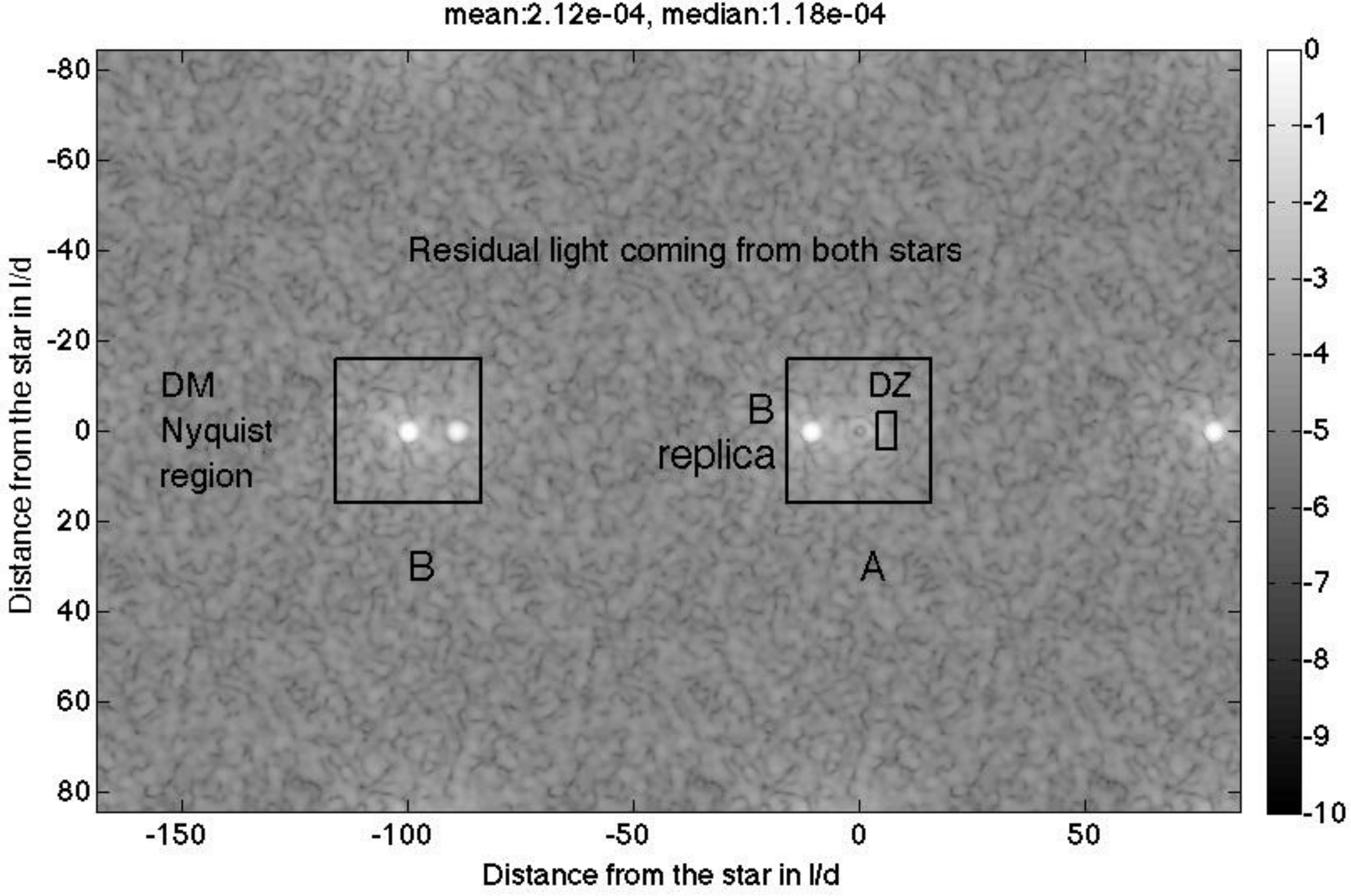}
 \caption{Diagram of Super-Nyquist Multi-Star Wavefront Control. The DM grating diffracts an attenuated replica of star B into a sub-Nyquist region of star
A, just as in the case of pure Super-Nyquist Wavefront Control. This brings the problem back to the simple Multi-Star Wavefront Control, for which we treat the diffracted replica as an actual star. In this image a coronagraph blocks the light originating from star A. A side effect that can be seen on the diagram is the replica of A in the controllable region of B. This allows us to then search for planets around A. }
\label{fig:SNMSWC} 
\end{center}
\end{figure}

In this section, we combined SNWC and MSWC and treat the case of two (or more) stars having a super-Nyquist separation. Figure \ref{fig:SNMSWC} shows a diagram of an on-axis star A (suppressed by a coronagraph if present), and the companion off-axis star B. Both stars have sub-Nyquist control regions where conventional WC can suppress the speckles field of one star but not the other. Suppose that the mild DM grating diffracts the (attenuated) replica of star B inside the sub-Nyquist region of star A, just as in the case of SNWC. As discussed in section \ref{sec:SNWC}, we can treat this diffracted replica as an actual star, effectively resulting in two stars with a sub-Nyquist separation. This reduces the problem to that of MSWC (which can be rigorously formalized at least in monochromatic light), which we demonstrated how to solve in section \ref{sec:MSWC}. The main potential difficulty arises for the case of broadband light, where the diffracted replica of the off-axis star looks spectrally elongated. This will certainly reduce the size of the correction region, but if SNWC works in broadband light then SNMSWC also works in principle with a reduced dark zone size.


\section{Simulation of SNWC}
\label{sec:simul}
To demonstrate the SNWC, we chose to simulate the observation of binary of equal brightness separated by 100 $\lambda/D$. This could be  the components of $\alpha$ Cen with a 1.5m telescope (such as Exo-S \footnote{http://exep.jpl.nasa.gov/stdt/exos/}) in monochromatic as well as polychromatic light (10\% bandwidth with a central wavelength of 770nm). The expected separation of $\alpha$ Cen components A and B in the year 2025 is about 10 arcsec, which corresponds to about 100 $\lambda/d$ away at a wavelength of 770 nm. This is also applicable to imaging large disks.
Since this paper demonstrates SNWC and not MSWC, we are assuming that the light from the target on-axis star has been completely removed (e.g. by a starshade) and only the off-axis companion remains. The goal is to demonstrate in simulation that it is indeed possible to create a dark zone beyond the Nyquist frequency of a DM in this configuration.

\subsection{Simulation description}
In this paper,  we have focused on demonstrating wavefront correction, assuming we have perfect knowledge of the phase and amplitude of each star's speckle field in the region of interest. 

The problem of correction is conceptually separate from the problem of estimation and is in some sense more fundamental, because the ability to correct implies estimation is possible, but  the converse does not hold.

Conventional WC estimation (e.g. in EFC or stroke minimization, \citep{Giveon07, Pueyo10}) uses the DM to create known Òprobe fieldsÓ to modulate the region of interest and then analyzes this modulation to reconstruct the phase and amplitude of the pre-existing speckle field. Similarly to the correction problem, the estimation problem in all
cases (SNWC, MSWC, SNMSWC) can also be reduced to conventional wavefront estimation as long as the DM is able to create known Òprobe fieldsÓ in the region of interest of sufficient energy, which will be the case as long as we demonstrate that correction is possible. 

 In section \ref{sec:SNWC}, we described that SNWC requires the DM to diffract light beyond the Nyquist region with, for example, the periodic print-through pattern left over from manufacturing. To create such a pattern, we used a grid of dots in the image plane, created from a mask of uniform intensity equal to 1 with a grid of 0.  
 The period of the grid is set such that there will be a diffracted (and attenuated) PSF copy next to the zone we would like to correct.
However, even with a fixed number of actuators and therefore a fixed frequency created, (multiple of the number of actuators (here 32)), we can cover all separation scenarios.
Another parameter that can be adjusted in order to control the performance is the width of the dark zone region. Indeed the bigger the region, the more DM stroke is needed to achieve deeper contrast. We used a 4$\times$8 $\lambda/d$ region for the demonstration, which we found being a good compromise between the performance and the discovery region. 
The simulations were done both in monochromatic light (770 nm) and in polychromatic light with a 10\% band. Finally, we also studied the case of a non-aberrated and aberrated wavefront.  The aberrations were introduced in the pupil plane as a power law with a coefficient equal to -2.

\subsection{Results}
\begin{figure}[h]
\begin{center}
\begin{tabular}{cc}
\includegraphics[scale=0.45]{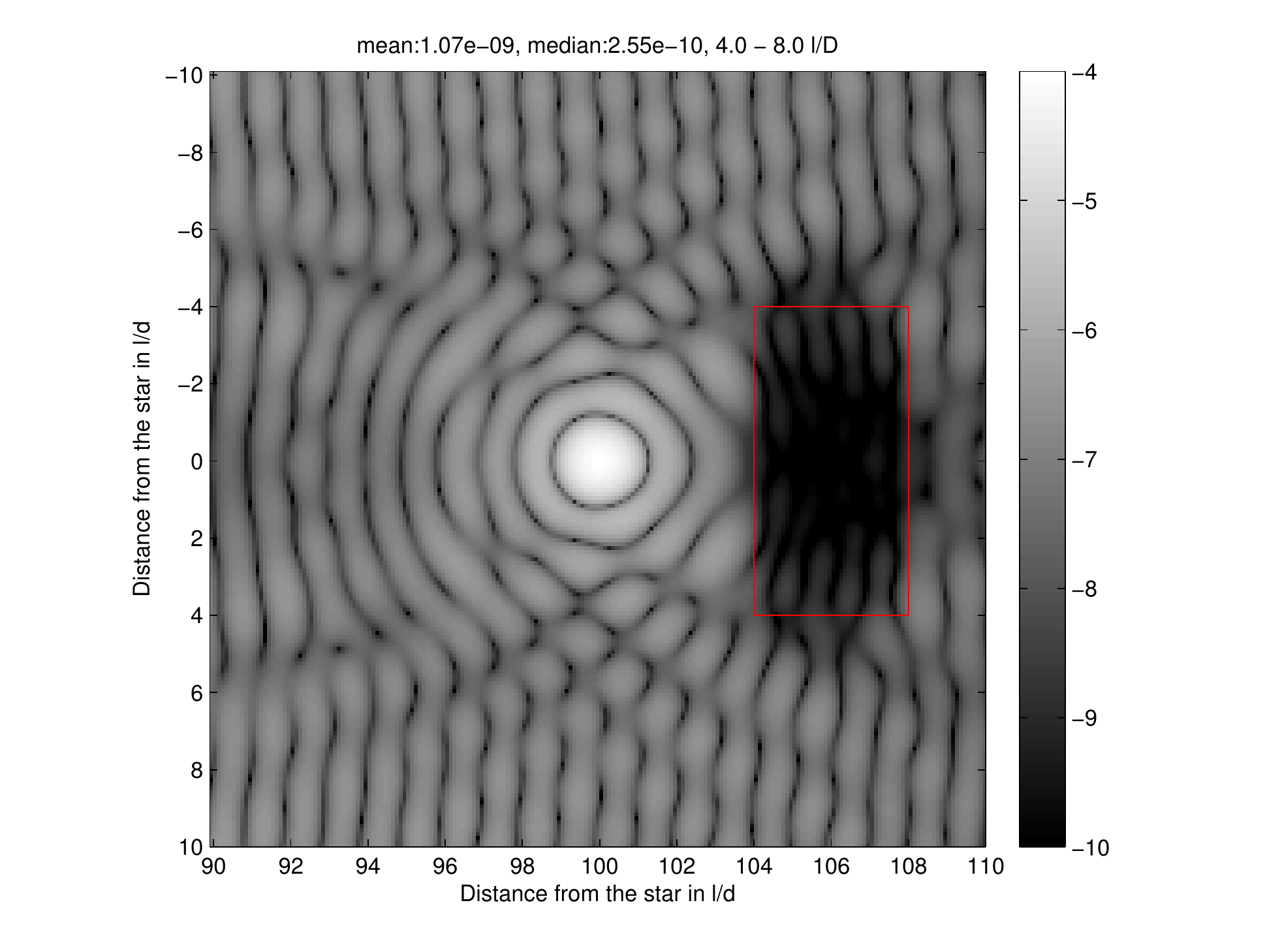}& \includegraphics[scale=0.45]{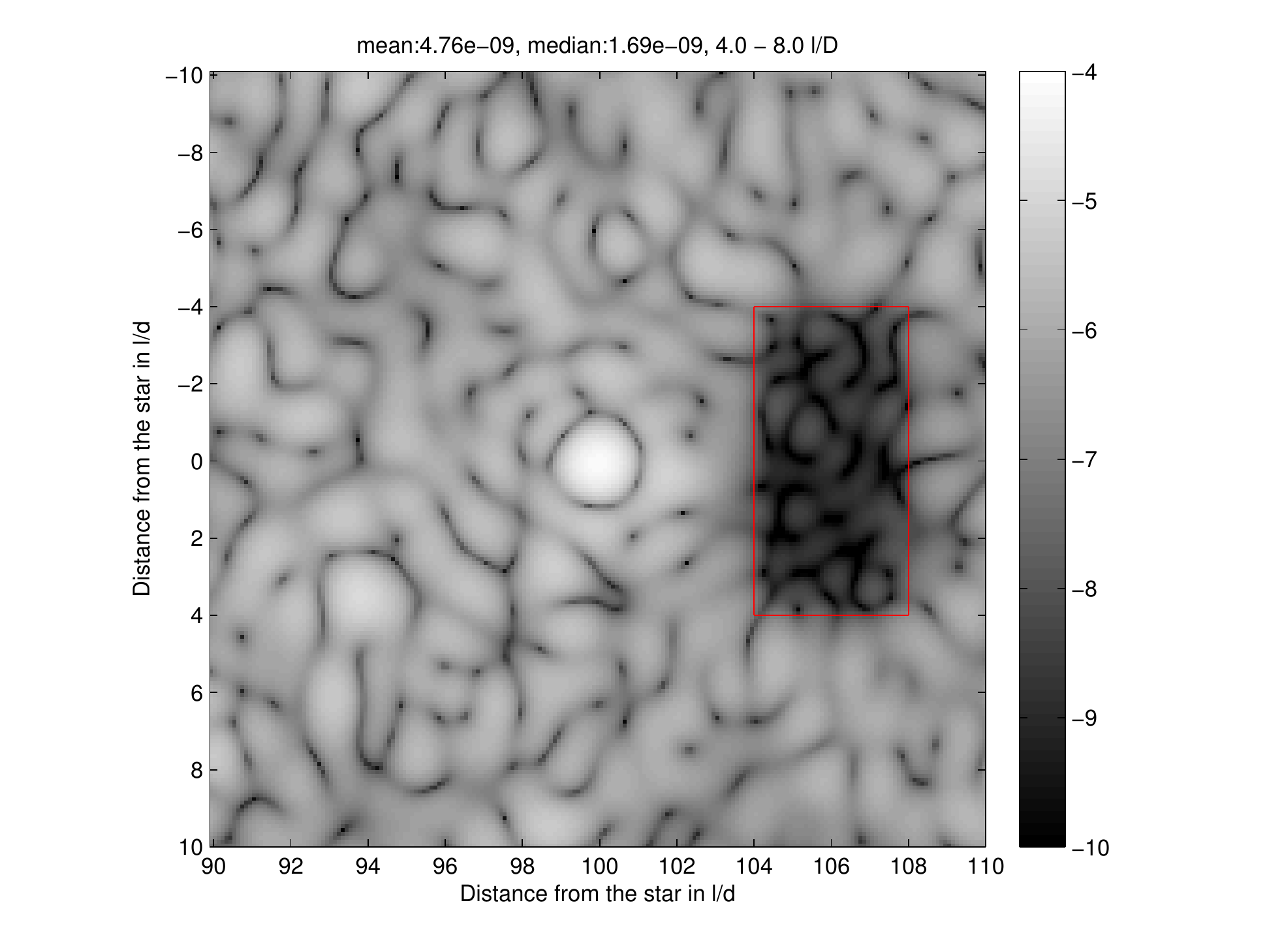}\\
\end{tabular}
\caption{Super-Nyquist simulation results in the monochromatic case, for a region of interest of 4x8 $\lambda/D$, Left: 0nm rms aberrations.  Right: 25nm rms aberrations. The median contrast obtained is  2.5$\times 10^{-10}$ without any aberrations and 1.7$\times 10^{-9}$ with aberrations.}
\label{fig:monolightres}
\end{center}
\end{figure}

\paragraph{Monochromatic Light}
In monochromatic light, the  introduced diffraction grid creates dots at 100 $\lambda$/d with an intensity of 1.34$\times 10^{-3}$ relative to the central star.  Figure \ref{fig:monolightres} shows the results of a (nonlinear) correction solution without aberrations (left) and with 25nm rms of aberrations. The median contrast obtained without any aberrations is 2.5$\times 10^{-10}$ and with aberration we reach a contrast of 1.7$\times 10^{-9}$. This corresponds to a factor 100 improvement from the no grid simulation for both the aberrated and non-aberrated case, demonstrating the ability of SNWC to create dark zones outside the nominal Nyquist limit of the DM. We expect that deeper contrasts are possible with better nonlinear solutions, but such optimizations are outside the scope of this paper.

\begin{figure}[h]
\begin{center}
\begin{tabular}{cc}
 \includegraphics[scale=0.45]{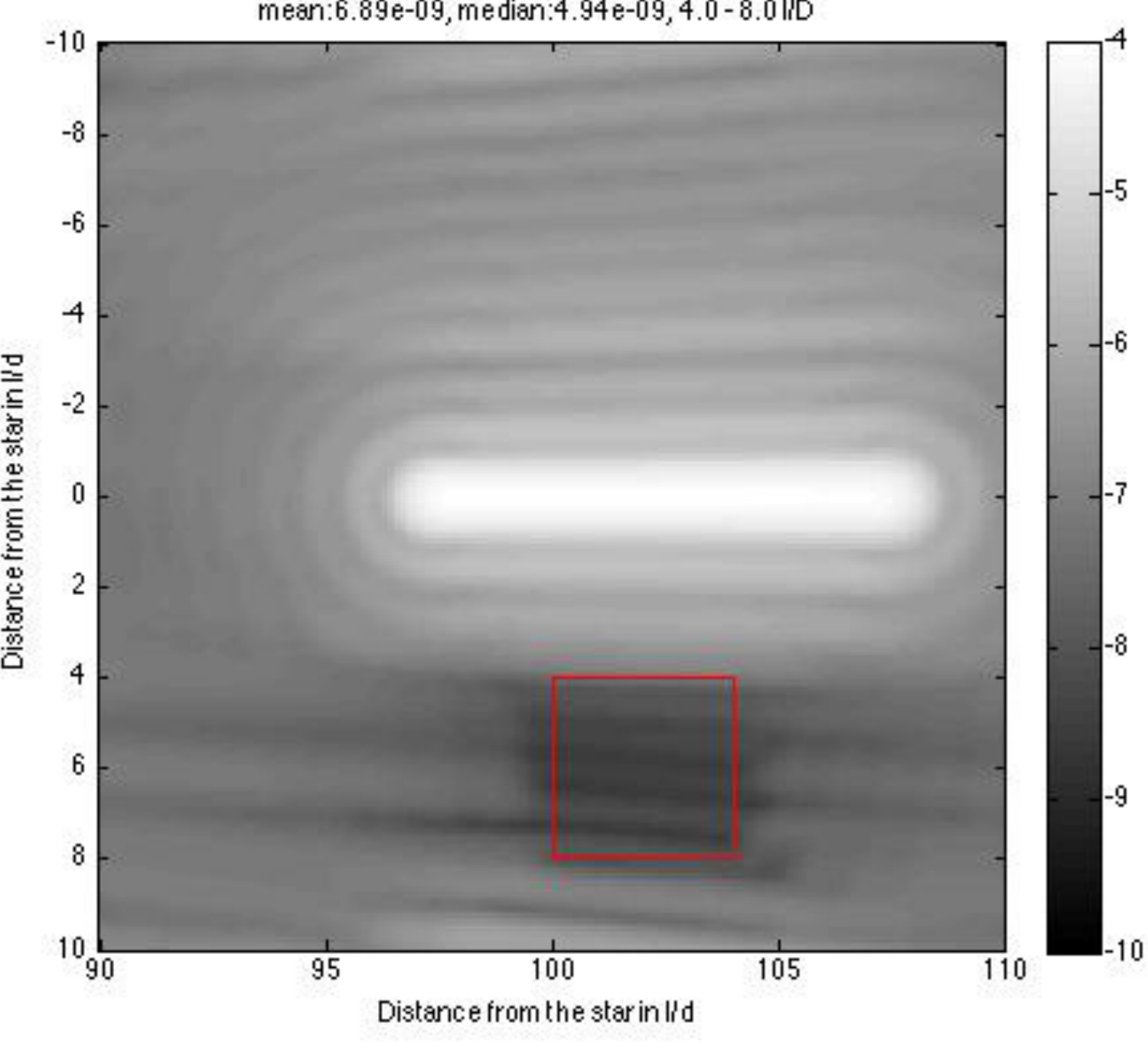}& \includegraphics[scale=0.45]{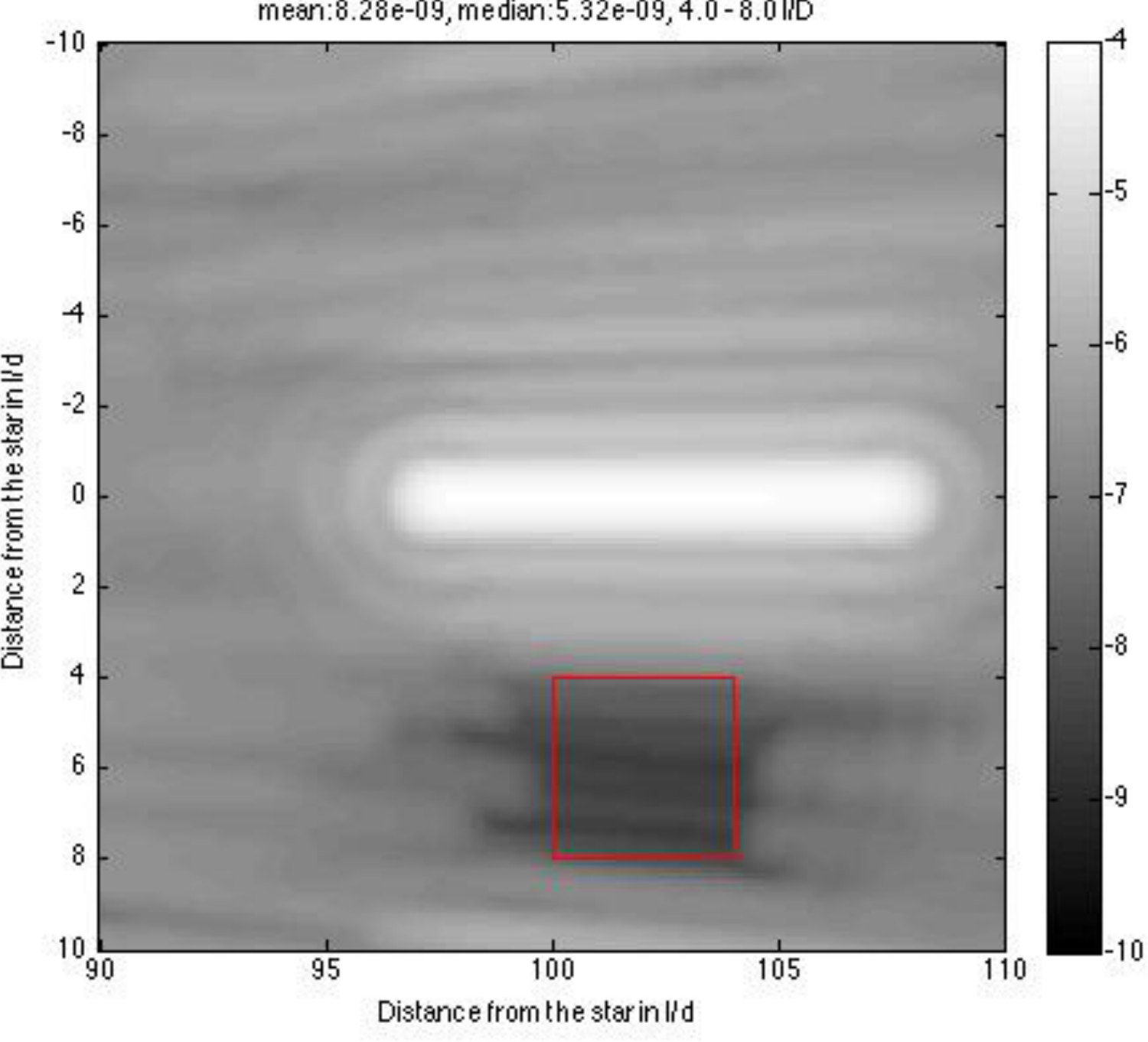}\\
\end{tabular}
\caption{ Super-Nyquist simulation results in the polychromatic case, 10\% bandwidth around 550nm. The dark zone region is around 4 and 8 $\lambda/D$. The figures on the left show the case of no aberrations and the figures on the right with 10nm rms of aberration. The median contrast obtained is 4.9x10-9 without aberrations and 5.3x10-9 with aberrations. This shows we are limited by polychromaticity.}
\label{fig:Poly10}
\end{center}
\end{figure}

\paragraph{Polychromatic Light}
We now consider a more realistic scenario and study the effect of polychromatic light. We use the methods in \citep{Giveon07} for polychromatic control within a 10\% bandwidth. In order to get an accurate image, we chose to sample the 10\%  bandwidth at 3 wavelengths. This allows a good compromise between computational speed and spectral resolution.  For better contrasts or larger bandwidth, one will need to increase the sampling of the bandwidth.
Figure \ref{fig:Poly10} shows the results without and with 10nm rms aberrations.  The median contrast obtained without any aberrations is 4.9$\times 10^{-9}$ and very similarly for the case with aberrations, we reach 5.3$\times 10^{-9}$. 
This gives the hint that we are limited by chromaticity. Keep in mind however that these results are raw contrasts before any type of post-processing techniques that would allow us to gain a factor 10 or even 100 on the contrast, which is encouraging for planet detection and characterization.  

Table \ref{tab:res} shows the summary of the different simulation results in the case of a equal brightness system.
\begin{table}[h]
\begin{center}
\caption{\label{tab:res} Summary of performance achieved with the simulations before and after wavefront control (WFC) and for monochromatic and polychromatic cases. }
\begin{tabular}{lcccccc}

 & Aberration & Contrast \\
\hline
No Grid,  Monochromatic, before WFC        & 0nm & 3.5 $10^{-8}$ \\
No Grid, Polychromatic(10\%), before WFC  & 0nm & 3.6 $10^{-8}$  \\
Grid, Monochromatic, after WFC                    &0nm&2.6 $10^{-10}$  \\
Grid, Polychromatic(10\%), after WFC              &0nm&4.9 $10^{-9}$\\
\hline
No Grid,  Monochromatic, before WFC       & 25 nm& 5.4 $10^{-7}$\\
No Grid, Polychromatic (10\%), before WFC & 10nm & 8.7 $10^{-7}$\\
Grid, Monochromatic, after WFC                      &25nm&4.8 $10^{-9}$\\
Grid, Polychromatic(10\%), after WFC               &10nm& 5.3 $10^{-9}$\\
\hline
\end{tabular}
\end{center}
\end{table}

This table shows raw contrast, before any post-processing algorithms. Therefore, it is very encouraging to see that it would be to detect Earth-like planets around both components of $\alpha$ Centauri.

\section{Conclusion}
In this paper, we presented a method that will enable high-contrast imaging of multiple stars systems. This new concept was broken in in three different steps, namely; (a) Super-Nyquist Wavefront Control (SNWC); (b) Multi-Star Wavefront Control (MSWC); (c) Super-Nyquist Multi-Star Wavefront Control (MSSNWC). We showed with simulations that it is possible to create a dark zone past the Nyquist frequency of a deformable mirror using a diffractive grid in the pupil. This grid can be the DM itself or an additional mask. Performance was shown in the case of $\alpha$ Cen, an interesting target since it is the closest potential earth-like planet hosts. This particular scenario can be applicable to a perfect star shade coronagraph for which we totally block the light coming from the parent star but still need to remove the light coming from the companion in the dark zone of interest.
The next step is to show that the DM can create a dark zone when adding both components in the lab. This will be done either using one or two DMs. The main scientific impact of this work is to enable direct imaging of planetary systems and disks around multiple star systems as well as in regions far from the star. This can be done at little additional hardware cost or changes to existing mission concepts, such as AFTA, Exo-C, Exo-S, and EXCEDE will greatly multiply the science yield of these missions. 

\acknowledgments
The material is based upon work supported by the National Aeronautics and Space Administration under Prime Contract Number NAS2-03144 awarded to the University of California, Santa Cruz, University Affiliated Research Center. This work was supported in part by the National Aeronautics and Space Administration's Ames Research Center under a center investment fund. It was carried out at the NASA Ames Research Center. Any opinions, findings, and conclusions or recommendations expressed in this article are those of the authors and do not necessarily reflect the views of the National Aeronautics and Space Administration.

\bibliography{AlphaCen_v8}   
\bibliographystyle{apj}
\end{document}